\documentclass{aa} 
\usepackage{graphicx}
\usepackage{txfonts}
\usepackage{color}

\begin{document}

   \title{Effect of gravitational stratification on the propagation of a CME}

\author{P.Pagano\inst{1} \and D.H. Mackay\inst{1} \and S.Poedts\inst{2}}
\institute{{School of Mathematics and Statistics, University of St Andrews, North Haugh, 
St Andrews, Fife, Scotland KY16 9SS, UK.}
\and
{Dept. of Mathematics, Centre for Mathematical Plasma Astrophysics, KU Leuven, Celestijnenlaan 200B, 3001 Leuven, Belgium.}
}
   \date{}

 
  \abstract
   {Coronal mass ejections (CMEs) are the most violent phenomenon found on the Sun. One model that explains their occurrence is
the flux rope ejection model. A magnetic flux rope is ejected from the solar corona and reaches the interplanetary space
where it interacts with the pre-existing magnetic fields and plasma.
Both gravity and the stratification of the corona affect the early evolution of the flux rope.}
   {Our aim is to study the role of gravitational stratification on the propagation of CMEs.
    In particular, we assess how it influences the speed and shape of CMEs and
    under what conditions the flux rope ejection becomes a CME or
    when it is quenched.}
   {We ran a set of MHD simulations that adopt an eruptive initial magnetic configuration that has already been shown to be suitable
    for a flux rope ejection.
    We varied the temperature of the backgroud corona and the intensity of the initial magnetic field to tune
    the gravitational stratification and the amount of ejected magnetic flux.
    We used an automatic technique to track the expansion and the propagation of the magnetic flux rope in the MHD simulations.
    From the analysis of the parameter space, we evaluate the role of gravitational stratification on the CME speed and expansion.}
   {Our study shows that gravitational stratification plays a significant role in determining whether
    the flux rope ejection will turn into a full CME or whether the magnetic flux rope will stop in the corona.
    The CME speed is affected by the background corona where it travels faster
    when the corona is colder and when the initial magnetic field is more intense.
    The fastest CME we reproduce in our parameter space travels at $\sim850$ $km/s$.
    Moreover, the background gravitational stratification plays a role in the side expansion of the CME, and we find that
    when the background temperature is higher, the resulting shape of the CME is flattened more.}
   {Our study shows that although the initiation mechanisms of the CME are purely magnetic,
    the background coronal plasma plays a key role in the CME propagation, 
    and full MHD models should be applied when one focusses especially
    on the production of a CME from a flux rope ejection.}


   \maketitle
%

\section{Introduction}

Coronal mass ejections (CMEs) are the main drivers of space weather,
a term used to describe the effect of plasmas and magnetic fields
on the near Earth environment.
Although the precise mechanism that causes a CME on the Sun is still unclear,
the ejection of a magnetic flux rope from the solar corona
successfully describes many of the general features of CMEs
(\citet{ForbesIsenberg1991}, \citet{Amari2000}, \citet{FanGibson2007}).
After the ejection of the magnetic flux rope, 
the CME propagates through the solar corona and into interplanetary space.
Understanding CME propagation is a key element in space weather.

A key characteristic of a CME is the three-part structure \citep{Hundhausen1987}. 
A CME is normally composed of a dense bow front,
followed by a low-density region and, finally,
a very dense core placed approximately at the centre of the curved front.
The propagation of this structure is normally used to infer 
the trajectory and speed of the CME.
Several studies have analysed
the kinematics of CMEs, and quoted
speeds span from 100 km/s to 3300 km/s,
with an average speed of about 500 km/s \citep{Gopalswamy2004}.
Some CMEs undergo an impulsive acceleration phase
followed by a propagation at constant speed \citep{Zhang2001}, 
while others are subject to relatively long acceleration \citep{ChenKrall2003}.
While CMEs may propagate at different speeds, all CMEs eventually couple with the solar wind once they reach a 
height of about $4$ $R_{\odot}$ \citep{Gopalswamy2000b}.
Similarly, CMEs expand while travelling outward (\citet{Ciaravella2001}, \citet{Lee2009}).

As CMEs propagate in the solar corona they interact with
the pre-exsiting plasma and magnetic fields.
The solar corona is a highly dynamic environment composed of magnetized plasma
whose global physical evolution can be described by magnetohydrodynamics (MHD) theory.
In the solar corona, plasma motions are primarily driven by the Lorentz force.
However, under equilibrium conditions we can assume a zero Lorentz force acting
on the plasma.
In such cases,
the plasma distribution in the corona can be described as being stratified by the effect of gravity.
The solar corona is inherently a multi-temperature environment,
although highly efficient thermal conduction tends to thermalize
the plasma, especially in the quiet corona.

The ratio between the thermal and magnetic pressure, the $\beta$ of the plasma
determines whether the local dynamics is governed
by the magnetic field ($\beta<1$) or the plasma ($\beta>1$).
\citet{Gary2001} describes the $\beta$ distribution in the corona
and shows that in the lower corona, $\beta<1$.
In contrast in the outer corona, the $\beta$ value has a more diversified distribution
where $\beta<1$ regions alternate with regions where $\beta>1$.
Moreover, when a CME occurs, it carries outward plasma and magnetic flux,
which compress the plasma and magnetic field ahead of it, thereby
changing the $\beta$ profile.
Despite this interaction, it is important to note that
the closed magnetic field of the ejected plasmoid allows neither the mixing nor thermalization of
the ejected plasma with the surrounding corona
(\citet{Ciaravella2001}, \citet{Pagano2007}).

Thermal pressure is generally considered to play
no role in the initial stages of the ejection of the flux rope,
but does become relevant
during the propagation  phase of the CME when the flux rope
travels at large radial distances.
Similarly, gravity tends to obstruct the ejection,
although it can significantly affect only fragments of the eruption.
\citet{Innes2012} observed fragments of the eruption falling back
to the Sun after an eruption.

Here we specifically address the role of gravitational stratification
on the CME propagation.
In our framework, the ejection is caused by an magnetic field configuration where the Lorentz force is not zero.
This magnetic field is added to the
stratified solar corona, which is initially in hydrostatic equilibrium
and decoupled from the magnetic field.
In particular, we start from the model of \citet{Pagano2013}
where an eruptive magnetic configuration
is used, and the full life span of a magnetic flux rope is described
by coupling the global non-linear force free model of 
\citet{MackayVanBallegooijen2006A} with MHD simulations
run with the AMRVAC code \citep{Keppens2012}.
In \citet{Pagano2013} a rather simple background density and thermal pressure profile
allowed a detailed study of the dynamics of the ejection.
Here, we extend that model
to include solar gravity and density stratification in the initial conditions.
We explore how the parameter space affects the eruption characteristics by tuning both
the temperature of the solar corona and the intensity of the magnetic field.

Several studies have included density stratification and gravity in CME modelling and propagation.
\citet{Pagano2007} study the role of the ambient magnetic field topology in the CME expansion and thermal insulation,
\citet{Zuccarello2009} focus on the CME initiation mechanisms including gravity and the solar wind.
\citet{Archontis2009} describe a flux rope ejection in the presence of gravity following a magnetic flux emergence event.
Finally, \citet{Manchester2004} model a CME considering the solar wind and gravity up to 1 AU, while
\citet{Reeves2010} simulate a solar eruption in 2.5D including solar gravity
and the effects of thermal conduction, radiation, and coronal heating.
Moreover, \citet{Roussev2012} study solar eruptions with a treatment of thermodynamics and an eruption initiated from below the corona, while the effect of solar wind heating has recently been studied by \citet{PomoellVainio2012}.

Although many observations have focussed on studying the properties of CMEs,
studies that simultaneously consider the density stratification, and the ejected plasma and magnetic flux are
difficult, owing the complexity of the different diagnostics involved.
Some recent work that describes the propagation of CMEs in the solar corona include
\citet{Gopalswamy2012} and \citet{Bemporad2011}.
Other studies give diverse examples of how the CME propagation 
can be modified according to the environment where it develops; e.g., \citet{Temmer2012} describe the interaction of a CME with another CME that preceded it.
Separately, some studies have been carried out to analyse the consequence of
the gravitational stratification of the solar corona (\citet{Guhathakurta1999}, \citet{Antonucci2004}, \citet{Verma2013}).
As observational studies are difficult to carry out, our simulations will shed some light on physical problems that are
still difficult to investigate from an observational view point.

The paper is structured as follows. In Sect.\ref{model}, we describe the numerical model,
in Sect.\ref{results} we describe the results of our simulations, in
Sect.\ref{discussion} we discuss the results, and the conclusions are drawn in Sect.\ref{conclusion}.

\section{Model}
\label{model}

To study the effect of gravitational stratification on the propagation of a CME,
we start from the work of \citet{Pagano2013}
where a magnetic configuration has been shown to be suitable for ejecting a magnetic flux rope. 
In the present work, we perform a number of changes to the modelling technique of \citet{Pagano2013},
in order to focus on the role of gravitational stratification.
We have carried out a set of MHD simulations that
consider variations in the temperature of the background
coronal atmosphere and the magnetic field intensity.

\subsection{MHD simulations}
We used the AMRVAC code developed at the KU Leuven to run the simulations \citep{Keppens2012}.
The code solves the MHD equations, and the terms
that account for gravity are included:
\begin{equation}
\label{mass}
\frac{\partial\rho}{\partial t}=-\vec{\nabla}\cdot(\rho\vec{v}),
\end{equation}
\begin{equation}
\label{momentum}
\frac{\partial\rho\vec{v}}{\partial t}+\vec{\nabla}\cdot(\rho\vec{v}\vec{v})=
   -\nabla p+\frac{(\vec{\nabla}\times\vec{B})\times\vec{B})}{4\pi}+\rho\vec{g},
\end{equation}
\begin{equation}
\label{induction}
\frac{\partial\vec{B}}{\partial t}=\vec{\nabla}\times(\vec{v}\times\vec{B}),
\end{equation}
\begin{equation}
\label{energy}
\frac{\partial e}{\partial t}+\vec{\nabla}\cdot[(e+p)\vec{v}]=\rho\vec{g}\cdot\vec{v},
\end{equation}
where $t$ is the time, $\rho$ the density, $\vec{v}$ velocity, $\vec{B}$ magnetic field, and
$p$ thermal pressure.
The total energy density $e$ is given by
\begin{equation}
\label{enercouple}
e=\frac{p}{\gamma-1}+\frac{1}{2}\rho\vec{v}^2+\frac{\vec{B}^2}{8\pi}
\end{equation}
where $\gamma=5/3$ denotes the ratio of specific heats.
The expression for the solar gravitational acceleration is
\begin{equation}
\label{solargravity}
\vec{g}=-\frac{G M_{\odot}}{r^2}\hat{r},
\end{equation}
where $G$ is the gravitational constant, and $M_{\odot}$ denotes the mass of the Sun.
In order to gain accuracy in the description of the thermal pressure,
we make use of the magnetic field splitting technique \citep{Powell1999},
as explained in Sect. 2.3 of \citet{Pagano2013}.

The initial magnetic field condition of the simulations is chosen such that it initially
produces the ejection of a flux rope and so the subsequent CME propagation can be studied.
The configuration of the magnetic field is
Day 19 in the simulation of \citet{MackayVanBallegooijen2006A}.
\citet{Pagano2013} in Sect. 2.2 explain in detail how the magnetic field distribution is imported from the
global non-linear force free field (GNLFFF) model of \citet{MackayVanBallegooijen2006A} to our MHD simulations.
Since the GNLFFF model contains no plasma, it needs to be defined.
We assume an initial atmosphere of a constant temperature corona stratified by solar gravity,
where we also allow for low background pressure and density:

\begin{equation}
\label{densitystratification}
\rho(r)=\rho_0 e^{{\frac{M_{\odot}G \mu m_p}{T k_b r}}}+\rho_{bg},
\end{equation}

\begin{equation}
\label{pressurestratification}
p(r)=\frac{k_b T}{\mu m_p}\rho_0 e^{{\frac{M_{\odot}G \mu m_p}{T k_b r}}}+p_{bg}
\end{equation}
where we use $\rho_0$ to tune the density at the lower boundary,
$\mu=1.31$ is the average particle mass in the solar corona,
$m_p$ is the proton mass,
$k_b$  Boltzmann constant, and
$T$ the temperature of the corona.
We tune $\rho_0$ depending on the temperature $T$ in order to always have the same density
at the bottom of our computational domain.
When $\rho_{bg}=p_{bg}=0$, we have a stratified atmosphere in hydrostatic equilibrium.
In some simulations, we need to have non-zero values for $\rho_{bg}$ and $p_{bg}$
 to avoid extremely low $\beta$ values at the outer radial boundary of the simulation.
These values are chosen such that $p_{bg}=\rho_{bg}/(\mu m_p) k_b T$.
In our simulations such a departure from equilibrium implies a negligible inflow of plasma
(in contrast to the real solar corona that is outflowing).
More details are given in the Appendix.
Finally, we set $\vec{v}=0$ as initial condition for the velocity.

Since the results of \citet{MackayVanBallegooijen2006A} are specified in terms of the potential vector $\vec{A}$,
we can uniformly multiply our initial magnetic field distribution
without hindering the validity of the results of \citet{MackayVanBallegooijen2006A}.
In the present work, we use the maximum value of the magnetic field intensity
of the initial magnetic field configuration, $B_{max}$, as a simulation parameter.
In our simulation, the magnetic field intensity is maxmium, $|B|=B_{max}$,
at the centre of the right-hand side bipole on the lower boundary.

The simulation domain extends over $3$ $R_{\odot}$ in the radial dimension starting from
$r=R_{\odot}$. The colatitude, $\theta$, spans from $\theta=30^{\circ}$ to
$\theta=100^{\circ}$ and the longitude, $\phi$, spans over $90^{\circ}$.
This domain extends to a larger radial distance than the domain used in \citet{MackayVanBallegooijen2006A}
from which we import the magnetic configuration.
To define the MHD quantities in the portion of the domain from $2.5$ $R_{\odot}$ to $4$ $R_{\odot}$,
we use Eqs.\ref{densitystratification} and \ref{pressurestratification} for density and thermal pressure
and the magnetic field for $r>2.5$ $R_{\odot}$ is assumed to be purely radial ($B_{\theta}=B_{\phi}=0$) where
the magnetic flux is assumed to be conserved
\begin{equation}
\label{brover25r}
B_r(r>2.5 R_{\odot},\theta,\phi)=B_r(2.5 R_{\odot},\theta,\phi)\frac{2.5^2}{r^2}.
\end{equation}

It should be noted that the initiation of the ejection is not
affected in any way by the extension of the magnetic field, since the initial dynamics
are a result of the flux rope that lies at about $1.2$ $R_{\odot}$ out of magnetohydrostatic equilibrium.

The boundary conditions are treated with a system of ghost cells.
Open boundary conditions are imposed at the outer boundary, reflective boundary conditions are set at the $\theta$ boundaries, and the $\phi$ boundaries are periodic.
The $\theta$ boundary condition is designed to not allow any plasma or magnetic flux through, while the $\phi$ boundary conditions allow the plasma and magnetic field to freely evolve across the boundaries.
These boundary conditions match those used in \citet{MackayVanBallegooijen2006A}.
In our simulations, the expanding and propagating flux rope only interacts with the $\theta$ and $\phi$ boundaries
near the end of the simulations, thus they do not affect our main results
regarding the initiation and propagation of the CME.
At the lower boundary, we impose constant boundary conditions taken from the first four $\theta$-$\phi$
planes of cells derived from the GNLFFF model.
The computational domain is composed of $256\times128\times128$ cells distributed in a uniform grid.
Full details of the grid can be found in Sect. 2.3 of \citet{Pagano2013}.

\subsection{Parameter space investigation}
\label{parameterspace}

\begin{figure}[!htcb]
\centering
\includegraphics[scale=0.60,clip,viewport=120 80 430 270]{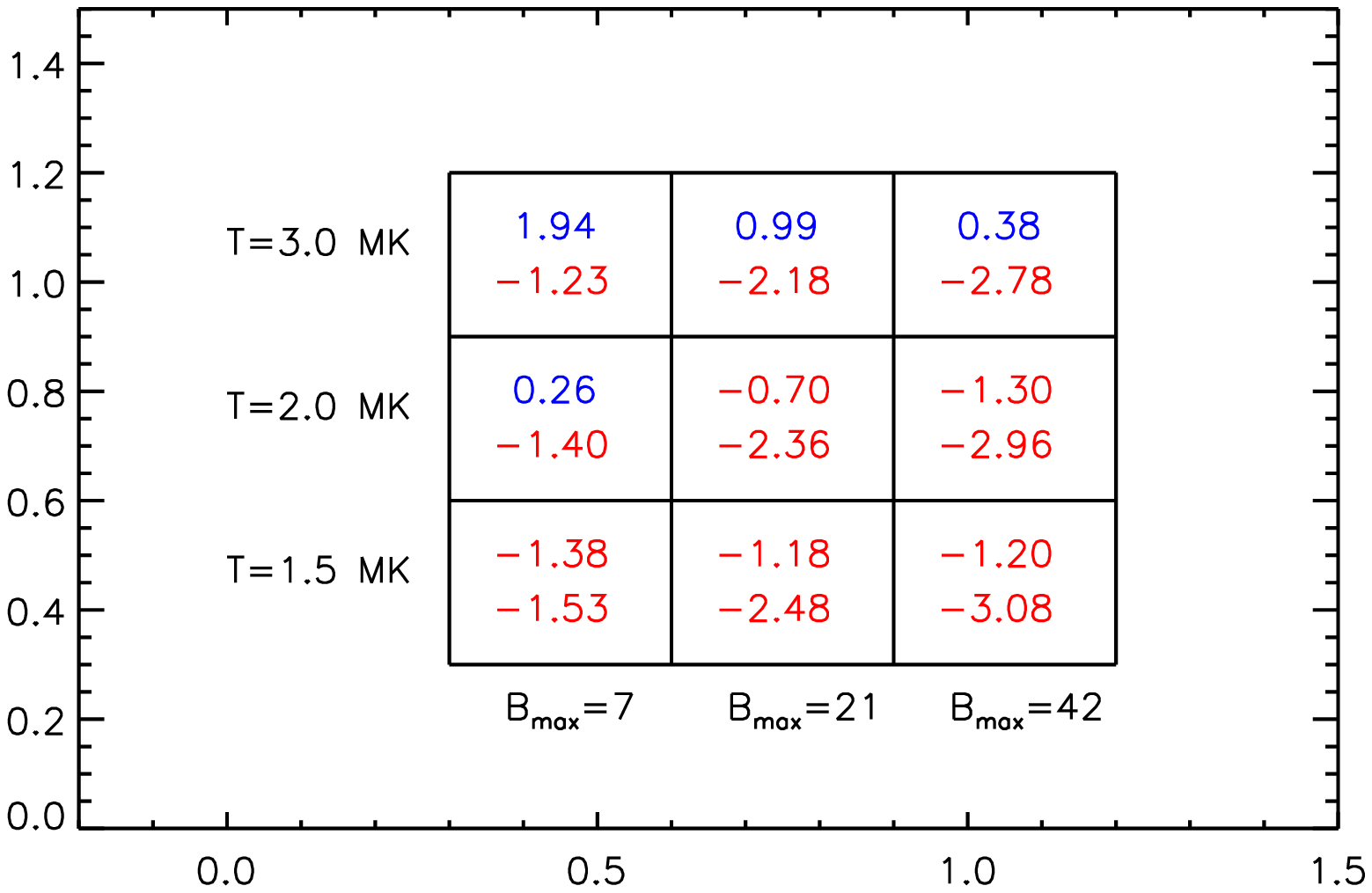}
\caption{Grid summarizing the parameter space we investigate. We ran 9 simulations with all the combinations of
$T=1.5$,$2$,$3$ $MK$ with $B_{max}=7$,$21$,$42$ $G$.
The numbers in the cell represent the $Log_{10}(\beta)$ at the lower boundary (lower number)
and the $Log_{10}(\beta)$ at the upper boundary (higher number). Red are negative values, and blue positive ones.}
\label{emptygrid}
\end{figure}

To analyse the role of the background stratified corona
we ran a set of nine simulations using three different temperatures ($T=1.5$,$2$,$3$ $MK$) of the corona
and three different maximum intensity values of the magnetic field ($B_{max}=7$,$21$,$42$ $G$), as illustrated in Fig.\ref{emptygrid}.

\begin{figure}[!htcb]
\centering
\includegraphics[scale=0.50,clip,viewport=25 10 490 340]{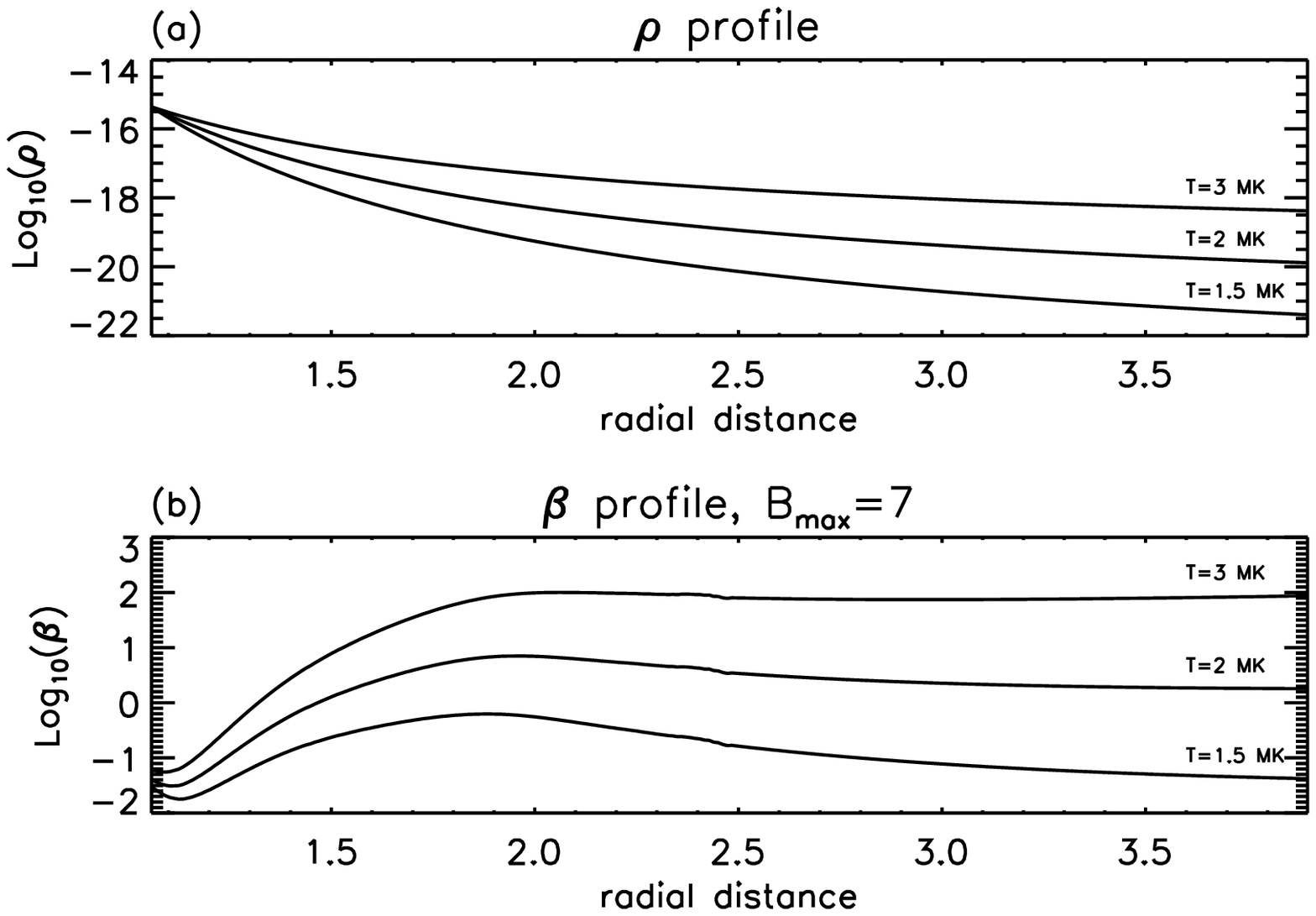}

\includegraphics[scale=0.50,clip,viewport=25 10 490 180]{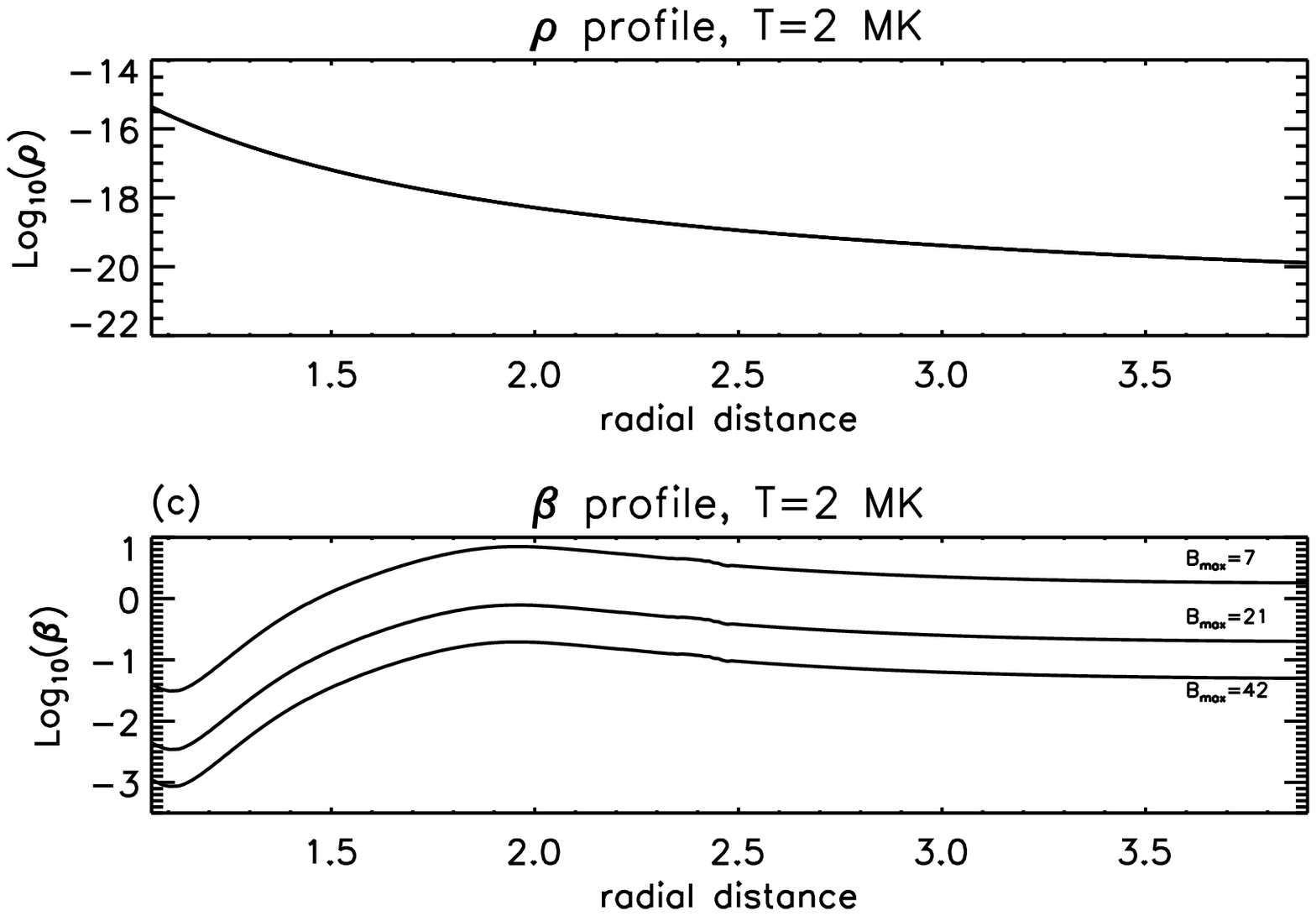}
\caption{(a) Profiles of $Log_{10}(\rho)$ along $r$ above the centre of the LHS bipole for
different temperatures of the solar corona, $T$.
(b) Profiles of $Log_{10}(\beta)$ along $r$ above the centre of the LHS bipole for
different temperatures of the solar corona, $T$, and with $B_{max}=7$ $G$.
(c) profiles of $Log_{10}(\beta)$ along $r$ above the centre of the LHS bipole with
different values of the parameter $B_{max}$ and with $T=2$ $MK$}
\label{profiles}
\end{figure}

In our model, a higher corona temperature implies a more uniform density and pressure gradient from the photosphere to the outer corona
and a higher amount of mass that constitutes the solar corona, as in Fig.\ref{profiles}a.
At the same time, the higher temperature leads to remarkably higher $\beta$ in the outer corona, as shown in Fig. \ref{profiles}b
for the $B_{max}=7$ $G$ case,
while in the lower corona ($r<1.2 R_{\odot}$) $\beta$ is clearly below $\sim10^{-1}$ regardless of the temperature.
For the $B_{max}=21$ and $42$ $G$ cases, the $\beta$ value is lower.
The outer corona switches from a low to a high $\beta$ regime when the temperature increases from $1.5$ to $3$ $MK$ (Fig.\ref{profiles}b).

We use the temperature to define a part of the parameter space since
it is the appropriate parameter to consistently tune
the profile of density and thermal pressure in the solar corona.
Since we always assume the same value for the photospheric density,
an increase in temperature implies a heavier column of mass placed above the magnetic flux rope.
By computing the integral of Eq.\ref{densitystratification} from $r=1.12$ $R_{\odot}$ to $r=4$ $R_{\odot}$,
we get a column of mass of $7.5 \times 10^{-6}$ $g/cm^2$ above the flux rope
for the simulation with $T=1.5$ $MK$ and $25$ times more for the simulation with $T=3$ $MK$.

At the same time, the pressure scale height reduces when lower temperatures are considered.
This implies that the ejected flux rope encounters less resistance from the compression produced ahead
when propagating.
The pressure length scale is around $0.06$ $R_{\odot}$ when the temperature is $T=1.5$ $MK$,
which doubles to $0.12$ $R_{\odot}$ when the temperature is $T=3$ $MK$.
It should be clarified, however, that in our simplified model for the gravitational stratification,
we do not aim to realistically describe the entire multi-temperature structure of the solar corona,
but rather the corona surrounding an active region.
In all of the simulations, the density and pressure drop steeply
in comparison to the full extent of the computational domain out to $4$ $R_{\odot}$.

Simulations with different $B_{max}$ (Fig.\ref{profiles}c) basically differ in the plasma $\beta$, which uniformly decreases as $B_{max}$ increases.
Thus, by changing the parameter $B_{max}$ we generally modify
the dominant forces and subsequently
the capacity of the solar corona to react to the flux rope ejection.

In some of the simulations ($T=1.5$ $MK$; $B_{max}=21$ and $B_{max}=42$ $G$), we use a non-zero $p_{bg}$ and $\rho_{bg}$
to avoid numerical problems due to extremely low $\beta$ at the external boundary.
For these simulations, the background pressure and density are four orders
of magnitude slower than the values near the lower boundary where the dynamics originate.
The resulting pressure and density profile departs from the analytic hydrostatic equilibrium profile, but we point out that
this can slow down the ejection only slightly,
because more plasma has to be displaced as the CME propagates due to the background density ($\rho_{bg}$).
However, the kinetic energy carried by the ejected plasma in the flux rope is much greater than the kinetic energy produced by these motions, and
these effects can in no way produce artificial ejections.
The tests described in the Appendix show that 
such departures from hydrostatic equilibrium only lead to appreciable effects on the magnetic configuration
on much larger time scales than the one for the CME to escape the solar corona in the present simulations.

\section{Results}
\label{results}

\begin{figure}[!htcb]
\centering
\includegraphics[scale=0.37,clip,viewport=120 40 900 630]{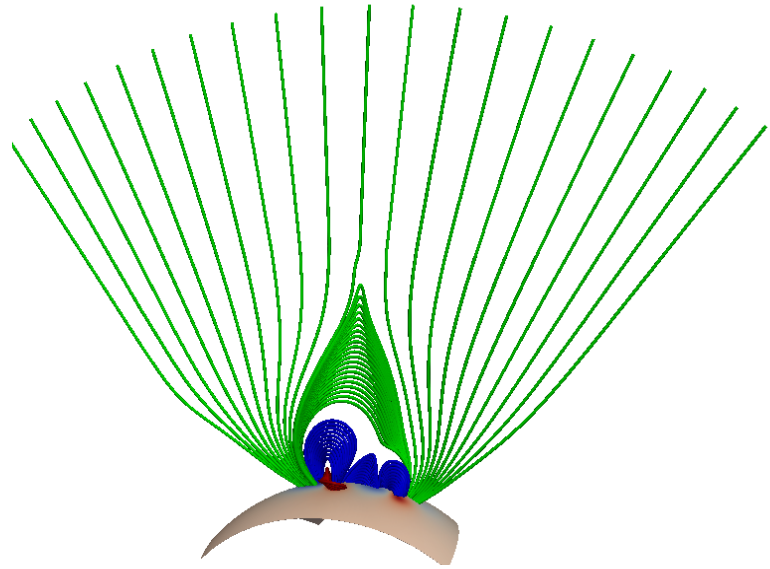}
\caption{Magnetic field configuration used as the initial condition in all the MHD simulations.
Red lines represent the flux rope, blue lines the arcades, green lines the external magnetic field.
The lower boundary is coloured according to the polarity of the magnetic field
from blue (negative) to red (positive) in arbitrary units.}
\label{initialmagnetic}
\end{figure}

\textcolor{green}{\textcolor[rgb]{0,0,0}{The initial magnetic field }}configuration is identical in all of the simulations and it
is chosen to produce the
ejection of the magnetic flux rope due to an initial excess of the Lorentz force.
Figure \ref{initialmagnetic} shows the initial magnetic configuration common to all the simulations.
The only difference from the one used in \citet{Pagano2013} is the larger extension of the domain in the radial direction.
The flux rope that is about to erupt lies under the arcade system,
and external magnetic field lines are shown above.
Some of the external magnetic field lines belong to the external arcade. while others are open.
A full description of the initial magnetic field configuration is given in Sect.3.1 of \citet{Pagano2013}.

In \citet{Pagano2013} the dynamics of the ejection and the production of the initial condition
in the GNLFFF simulation is discussed in detail.
Here, we only focus on the propagation of the flux rope once the ejection is triggered.

We first describe the characteristics of a typical ejection in this framework.
Following this we compare some key features between different simulations
in order to highlight the role of the temperature of the stratified corona ($T$)
and the intensity of the magnetic field ($B_{max}$).

\subsection{Typical simulation, ($T=2 MK$, $B_{max}=21$ $G$)}
\label{typicalsimulation}
\begin{figure}[!htcb]
\centering
\includegraphics[scale=0.17,clip,viewport=260 20 740 460]{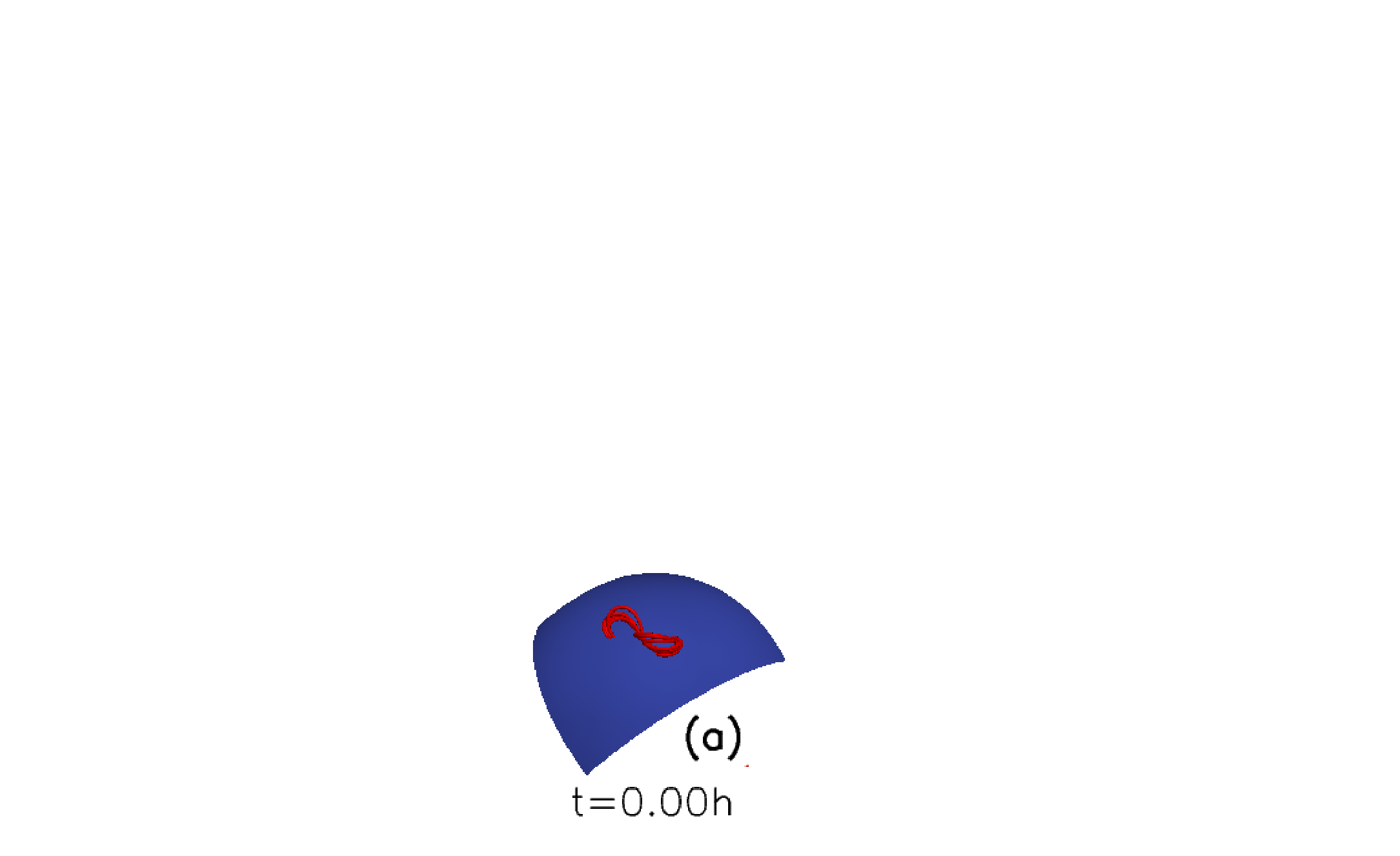}
\includegraphics[scale=0.17,clip,viewport=260 20 740 460]{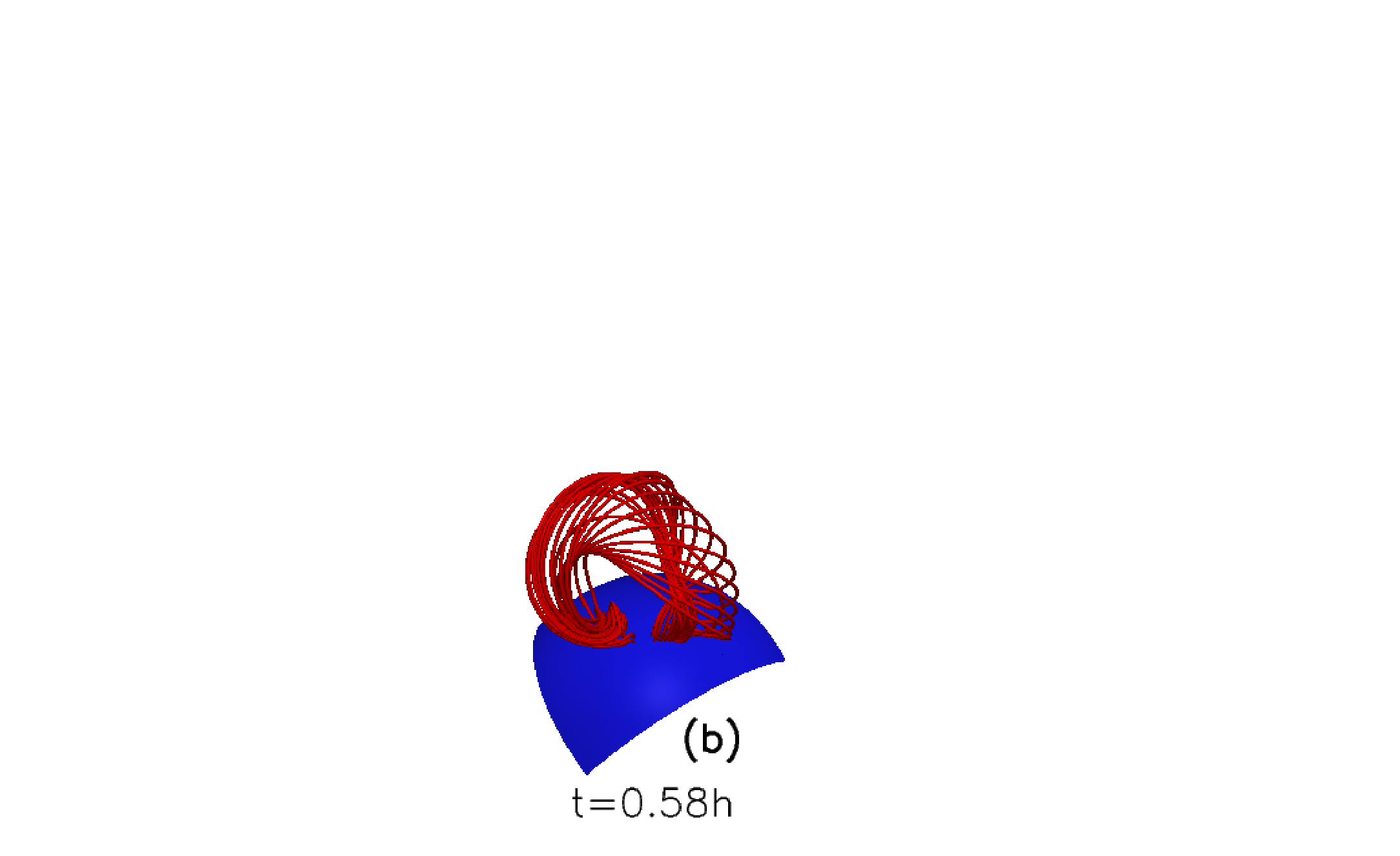}
\includegraphics[scale=0.17,clip,viewport=260 20 740 460]{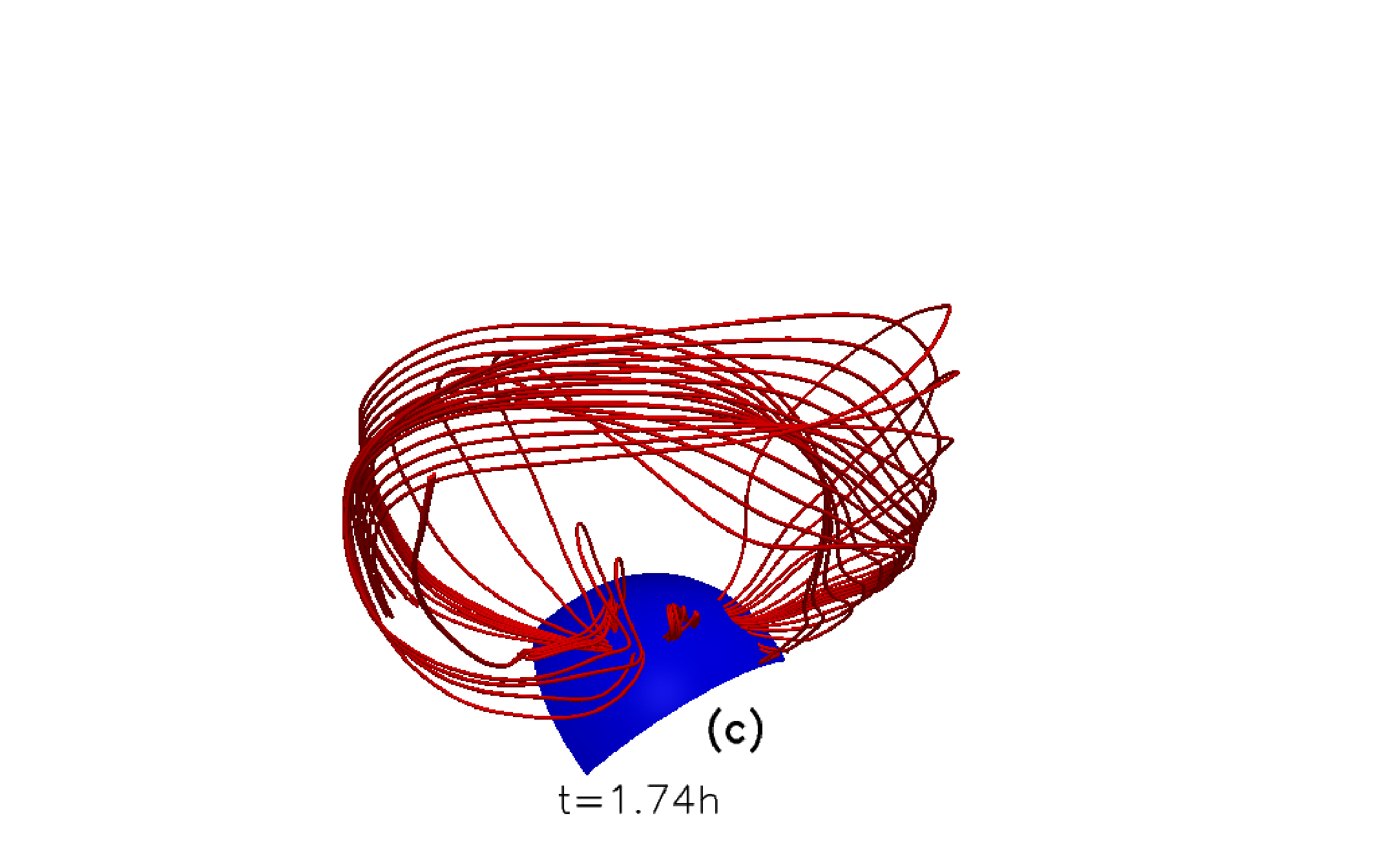}
\caption{The ejection of the magnetic flux rope. The red lines are all magnetic flux rope lines draw from both
the footpoints of the flux rope and from the centre of its axis.}
\label{fluxropeejection}
\end{figure}

A typical simulation is one with $T=2$ $MK$ and $B_{max}=21$ $G$, 
which is the central one in the parameter space grid shown in Fig.\ref{emptygrid}.
In this simulation the flux rope escapes the computational domain at $4$ $R_{\odot}$
and a CME occurs.
In Fig.\ref{fluxropeejection} we show the ejection and expansion of the flux rope in 3D to illustrate the simulation
with some of the magnetic field lines drawn from the flux rope footpoints and from the axis of the flux rope at different times.
In future figures we draw cuts in 2D planes to focus more clearly on specific aspects.

In the initial magnetic configuration (Fig.\ref{initialmagnetic}),
the flux rope lies above the centre of the left-hand side (LHS) bipole, three magnetic arcades connect
adjacent and opposite polarities, and a larger arcade connects the opposite external polarities of the two bipoles.
The flux rope that is in non-equilibirium lies in the asymmetric part of the global configuration.
The density profile falls off with radial distance where
$\rho\sim10^{-15}$ $g/cm^{3}$ at the lower boundary and $\rho\sim10^{-20}$ $g/cm^{3}$ at the top boundary (Fig.\ref{profiles}a).
The plasma $\beta$ is approximately $10^{-2}$ near the flux rope and it increases to $10^{-1}$ radially above it (Fig.\ref{profiles}c).
We note that there are regions where $\beta$ reaches as much as $10^2$ in the corona (such as near null points),
where the magnetic field is weak.

\begin{figure}[!htcb]
\centering
\includegraphics[scale=0.21,clip,viewport=27 20 530 325]{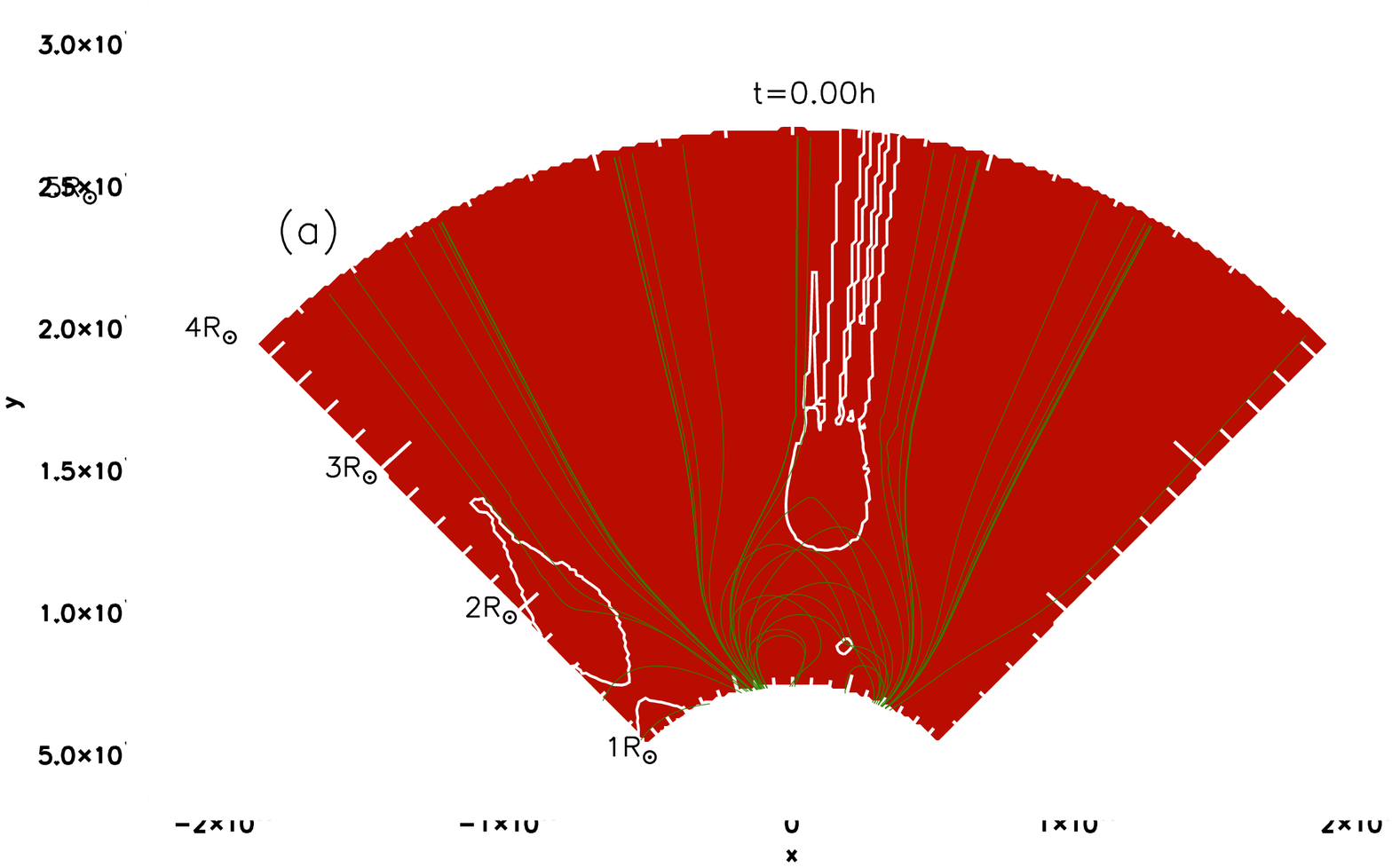}
\includegraphics[scale=0.18,clip,viewport=580 5 657 375]{Fig5a.ps}
\includegraphics[scale=0.21,clip,viewport=27 20 530 325]{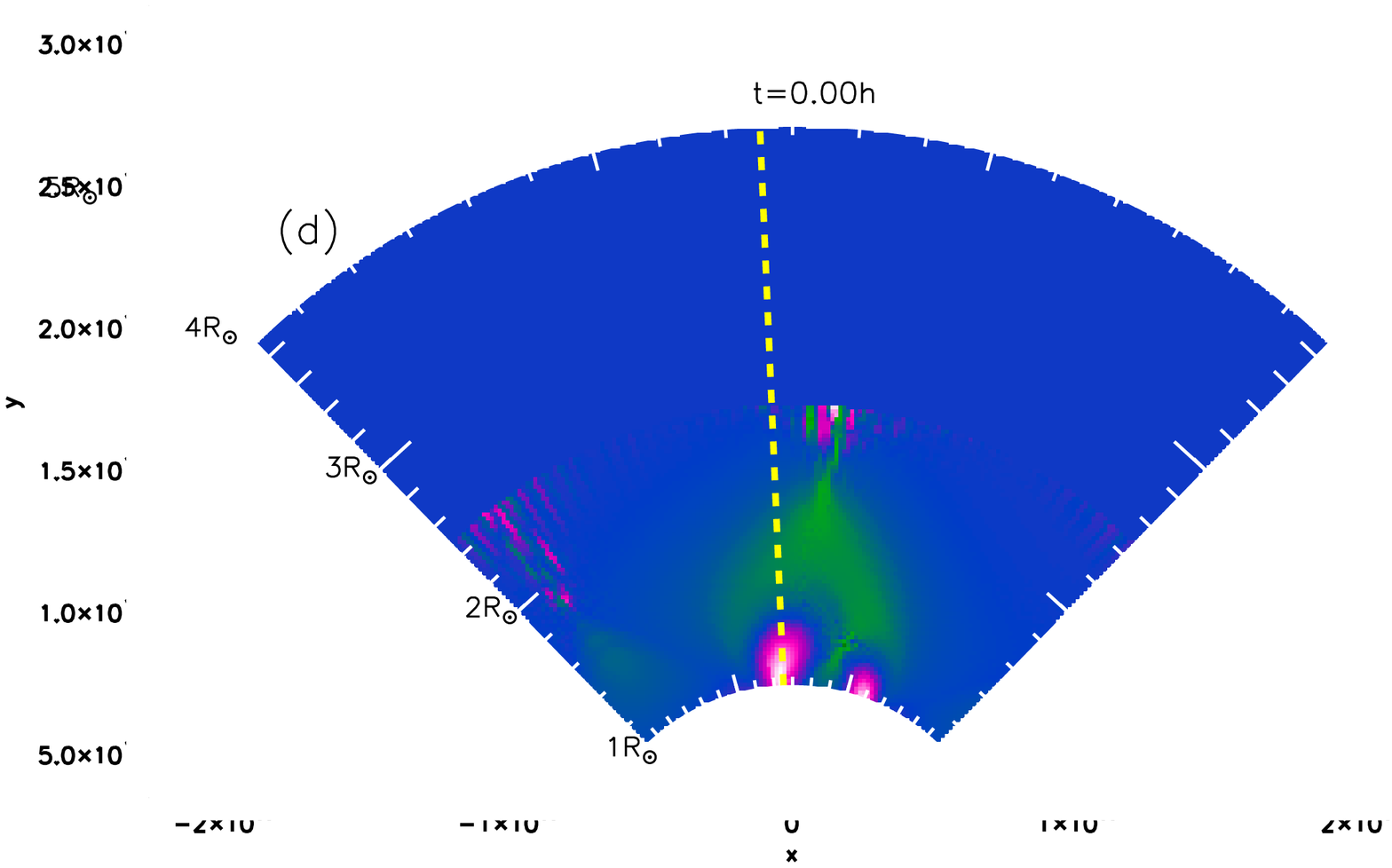}
\includegraphics[scale=0.18,clip,viewport=580 5 657 375]{Fig5d.ps}

\includegraphics[scale=0.21,clip,viewport=27 20 530 325]{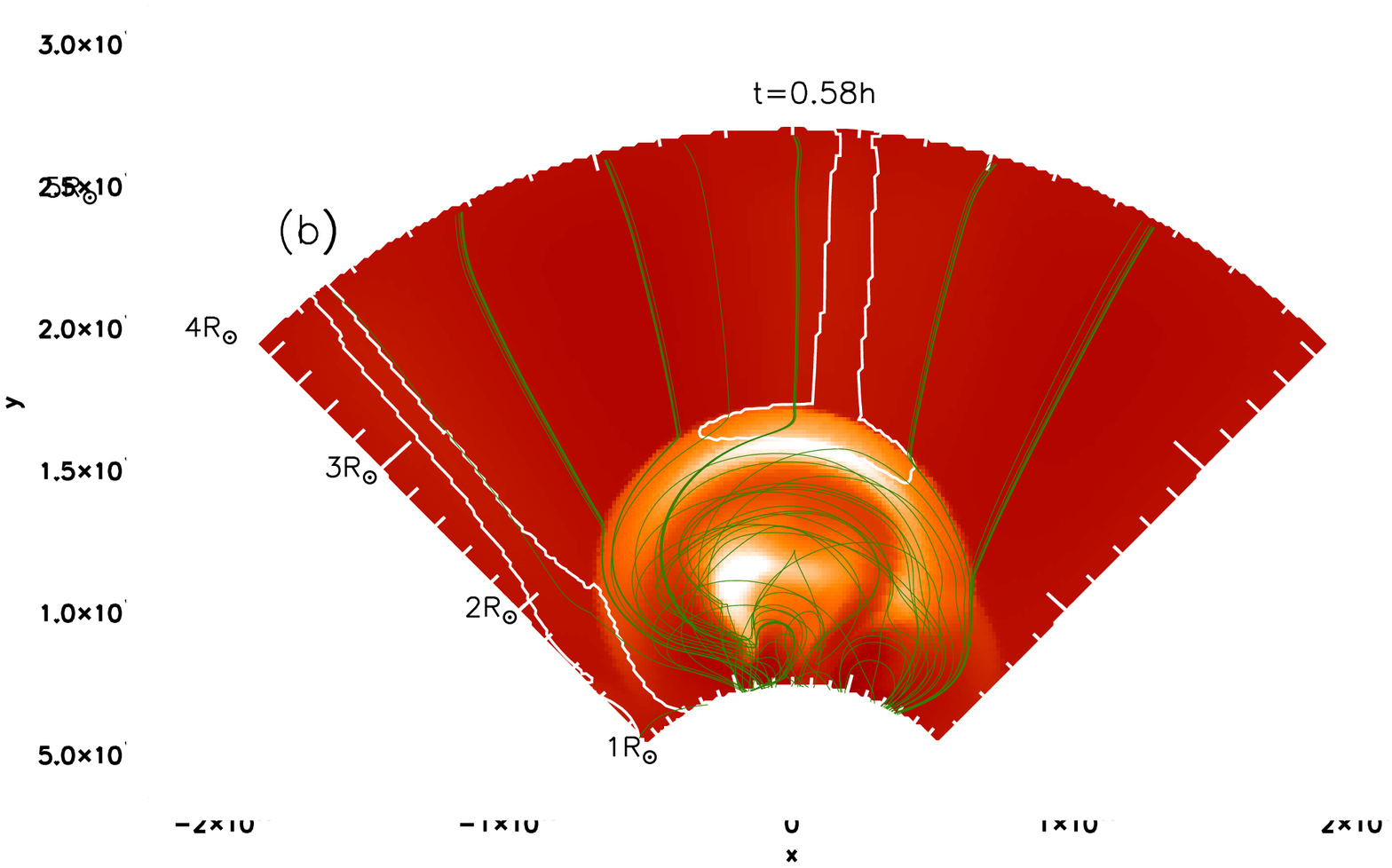}
\includegraphics[scale=0.18,clip,viewport=580 5 657 375]{Fig5b.ps}
\includegraphics[scale=0.21,clip,viewport=27 20 530 325]{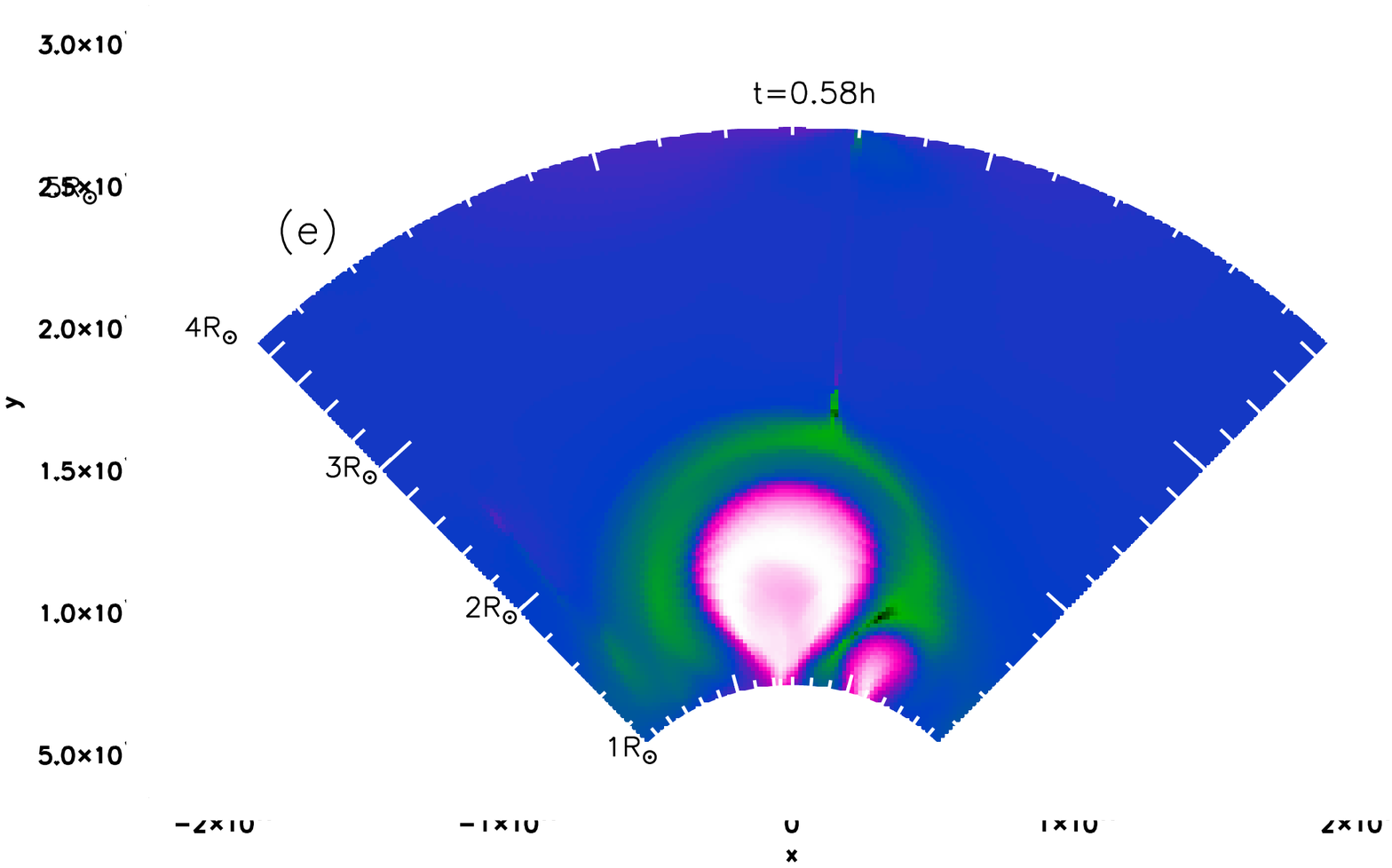}
\includegraphics[scale=0.18,clip,viewport=580 5 657 375]{Fig5e.ps}

\includegraphics[scale=0.21,clip,viewport=27 20 530 325]{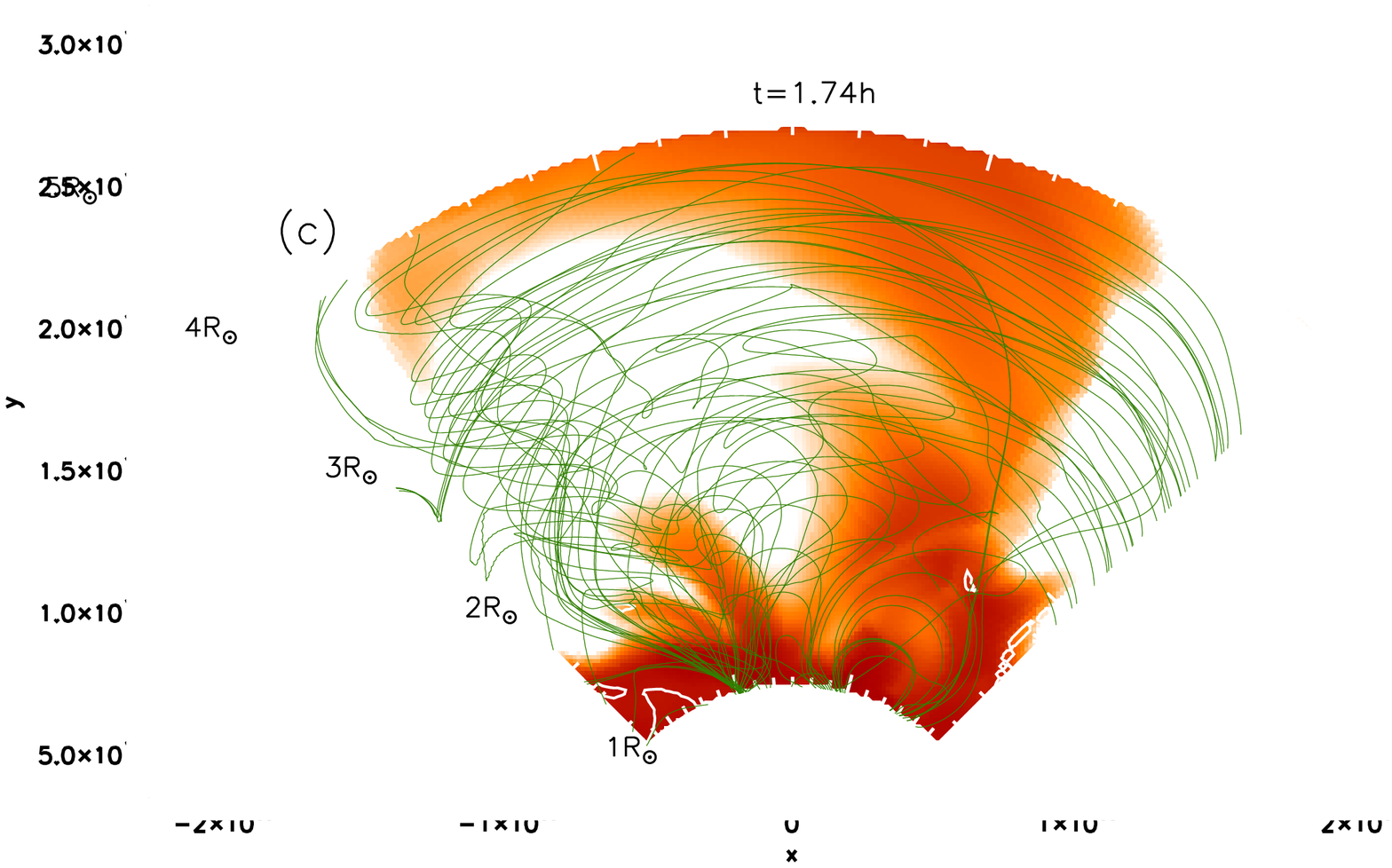}
\includegraphics[scale=0.18,clip,viewport=580 5 657 375]{Fig5c.ps}
\includegraphics[scale=0.21,clip,viewport=27 20 530 325]{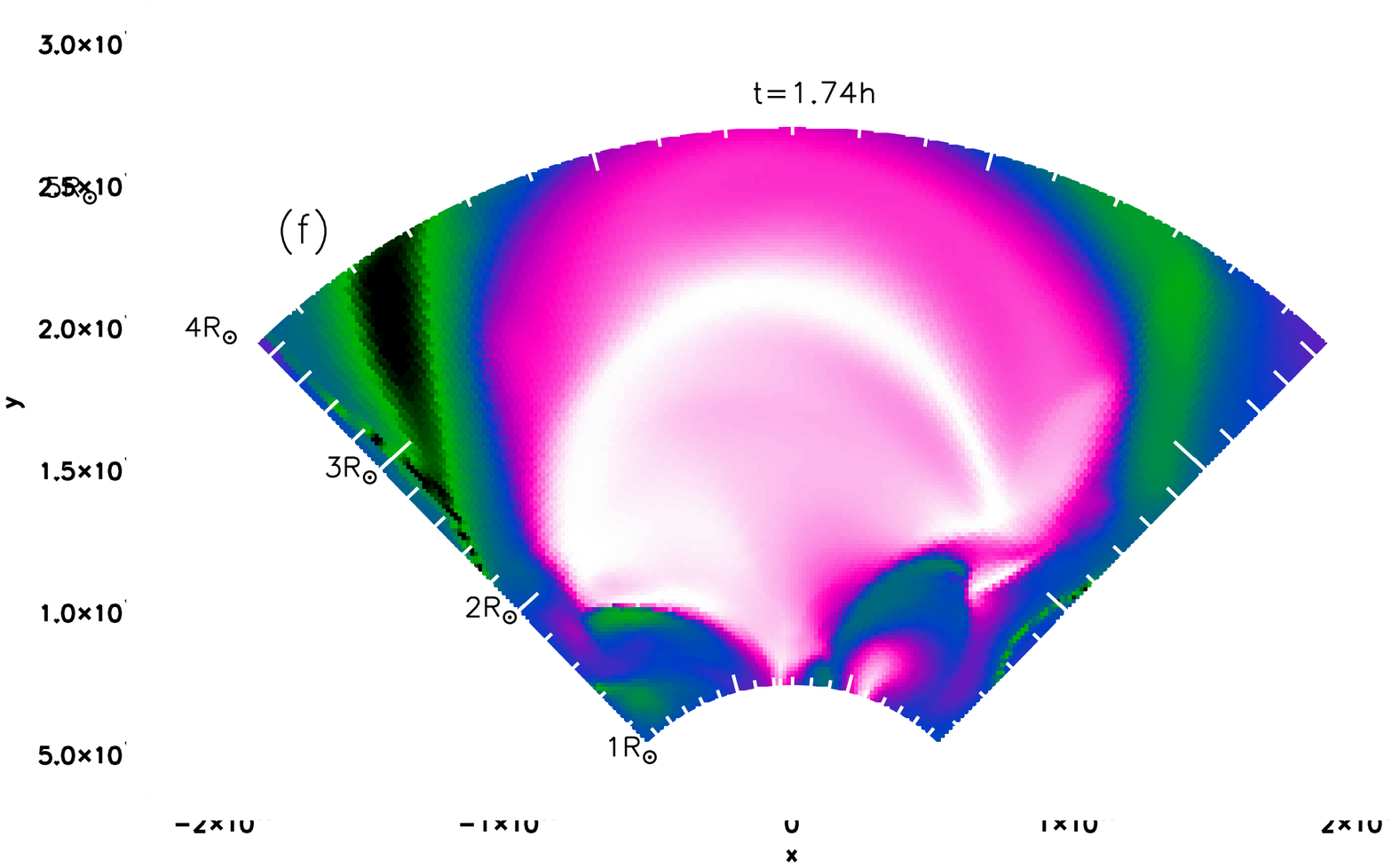}
\includegraphics[scale=0.18,clip,viewport=580 5 657 375]{Fig5f.ps}
\caption{Simulation with $T=2$ $MK$ and $B_{max}=21$ $G$.
(a)-(c) Maps of density contrast (Eq.\ref{densitycontrast}) in the $(r-\phi)$ plane
passing through the centre of the bipoles at $t=0$h, $t=0.58$h, $t=1.74$h.
Superimposed are magnetic field lines plotted from the same starting points (green lines)
and the contour line of $\beta=1$ (white line).
(d)-(f) Maps of $B_\theta/|B|$ on the same plane and at the same times.
Maps show the full domain of our simulations from $r=1$ $R_{\odot}$ to $r=4$ $R_{\odot}$.
In panel (d) the yellow dashed line is the cut for the plots in Fig.\ref{byfollowT2B21}.
The temporal evolution is available in the on-line edition.}
\label{evolT2B21}
\end{figure}

To follow the evolution of the simulation, in Fig.\ref{evolT2B21} we show two quantities in the $r-\phi$ plane
passing through the centre of the bipoles.
Figures \ref{evolT2B21}a-\ref{evolT2B21}c show the density contrast,
\begin{equation}
\rho_c=\frac{\rho(t)-\rho(t=0)}{\rho(t=0)}
\label{densitycontrast},
\end{equation}
that indicates the variation in density with respect to the initial condition, which is useful for marking the ejection of plasma.
Figsures \ref{evolT2B21}d-\ref{evolT2B21}f show the ratio $B_{\theta}/|B|$ that illustrates the evolution of the axial magnetic field of the flux rope,
which initially lies in the $\theta$-direction.
As explained in \citet{MackayVanBallegooijen2006A},
the formation of the flux rope due to magnetic diffusion
leads to a strong axial component of the magnetic field along the PIL of the LHS bipole.
Simultaneously, the increased tilt of the bipole leads to a quasi-antiparallel magnetic field
around the magnetic flux rope.
Because of the construction of the system of a E-W bipole and N-S PIL,
at $t=0$ the axial magnetic field of the flux rope is mostly along the $\theta$ direction
and the blue region between pink and green
in Figs. \ref{evolT2B21}d-\ref{evolT2B21}f, across which $B_{\theta}/|B|$
changes sign, and then marks the borders of the magnetic flux rope and the overlying arcade.
As long as no major reconnection occurs, the quantity $B_{\theta}/|B|$ is clearly positive above the bipoles
where the magnetic flux rope propagates and expands.

In this simulation, the flux rope is ejected outwards, and it leads
to an increase in density at larger radii (Fig.\ref{evolT2B21}a-Fig.\ref{evolT2B21}c),
and the propagation and expansion of the region 
where the magnetic field is mostly axial, i.e. $B_{\theta}/|B|\sim1$ (Figs. \ref{evolT2B21}c-\ref{evolT2B21}f).
The shape of the density propagation roughly reproduces the typical
three-component structure of a CME, which is clearly visible in Fig. \ref{evolT2B21}b.
A higher density bow front propagates upwards, and
behind it lies a region with lower density.
Finally, a dense core is located at the centre of the expanding dome.
In our simulation, the region with highest density coincides with the region where
$B_{\theta}/|B|$ is positive.
The quantity $B_{\theta}/|B|$ is maximum just ahead of the high- density region,
while the peak of density is highest roughly at the centre of the expanding structure
where  $B_{\theta}/|B|$ remains positive (Fig. \ref{evolT2B21}b).
This suggests that the magnetic flux rope is propagating upwards, perpendicular to its axial magnetic field lines,
lifting coronal plasma to produce the high density region.
The same process is seen in \citet{Pagano2013}, and it is
considered a standard process in filament eruptions.

\begin{figure}[!htcb]
\centering
\includegraphics[scale=0.50,clip,viewport=10 10 500 240]{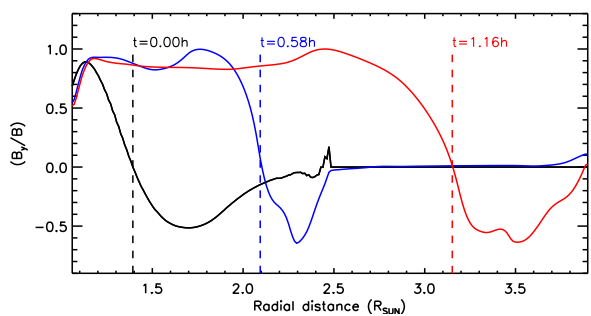}
\caption{Profile of $B_{\theta}/|B|$ above the centre of the LHS bipole, at different times (different colours).
Dashed lines of different colours indicate where $B_{\theta}/|B|=0$
and where we locate the top of the flux rope at a given time.}
\label{byfollowT2B21}
\end{figure}

At the same time, the region where a positive axial component of the magnetic field is dominant
expands and propagates upwards, roughly covering the same volume of space as that of the high-density region
(white and pink region in Figs. \ref{evolT2B21}d-\ref{evolT2B21}f).
Since the region where $B_{\theta}/|B|>0$ roughly reproduces the
high- density region that corresponds to the CME,
we use this quantity to track the ejection and expansion of the CME.
Figure \ref{byfollowT2B21} shows the profile of $B_{\theta}/|B|$ radially from the centre of the LHS bipole at different times.
The radial position where $B_{\theta}/|B|=0$ 
along the radial direction vertically from the centre of the LHS bipole is defined to be
the top of the magnetic flux rope in our representation.

In the plot in Fig. \ref{byfollowT2B21} we see the top of the flux rope at $1.4$ $R_{\odot}$
at $t=0$h (black line), and it reaches $3.15$ $R_{\odot}$ after $1.16$h (red line).
This clearly shows the ejection of the flux rope.

\subsection{Parameter space investigation}
\label{spaceparamaterCME}

\subsubsection{Quenched ejections}
\label{quenchedejections}
\begin{figure}[!htcb]
\centering
\includegraphics[scale=0.33,clip,viewport=27 20 530 325]{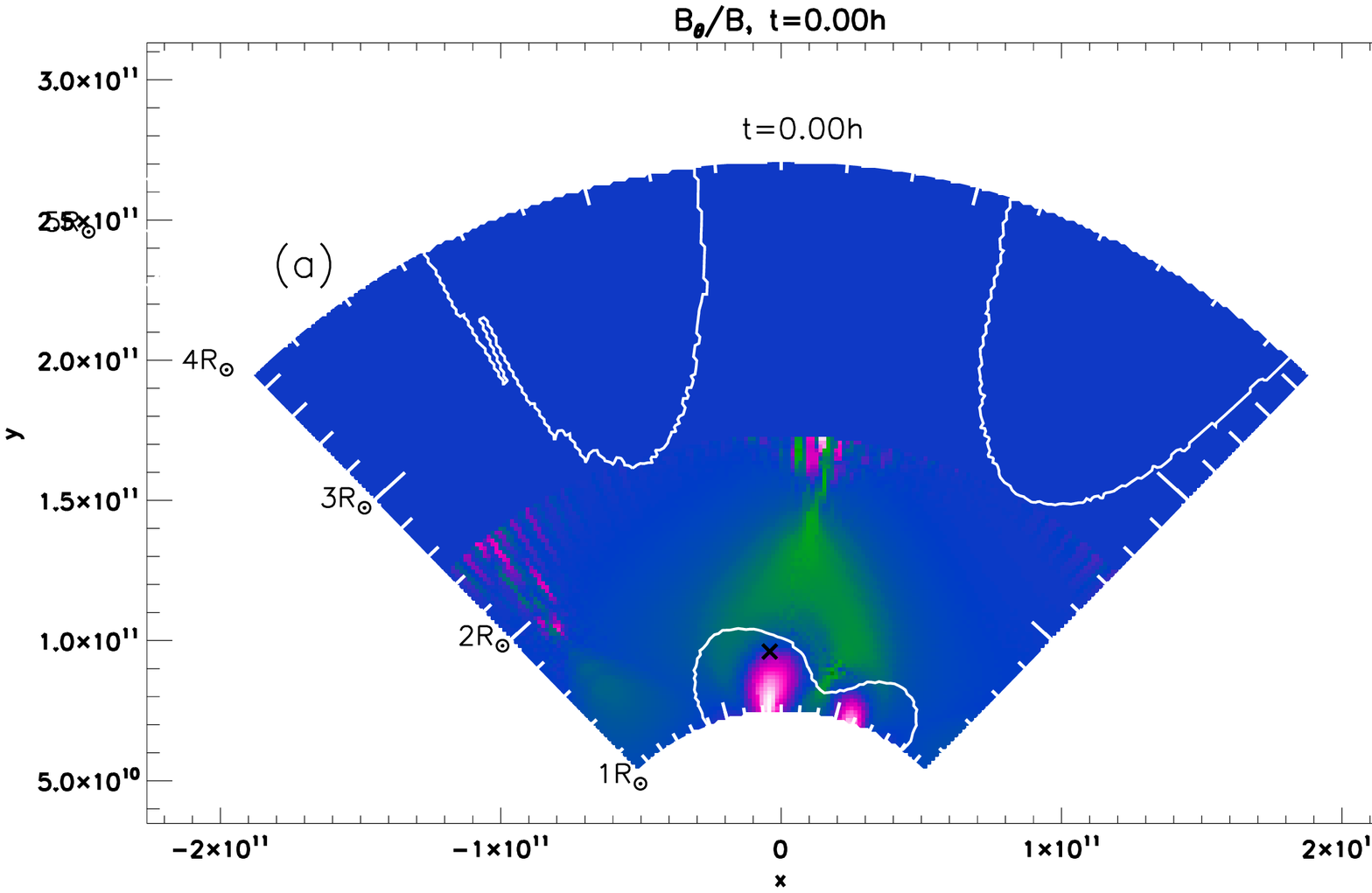}
\includegraphics[scale=0.28,clip,viewport=580 5 657 375]{Fig7a.ps}
\includegraphics[scale=0.33,clip,viewport=27 20 530 325]{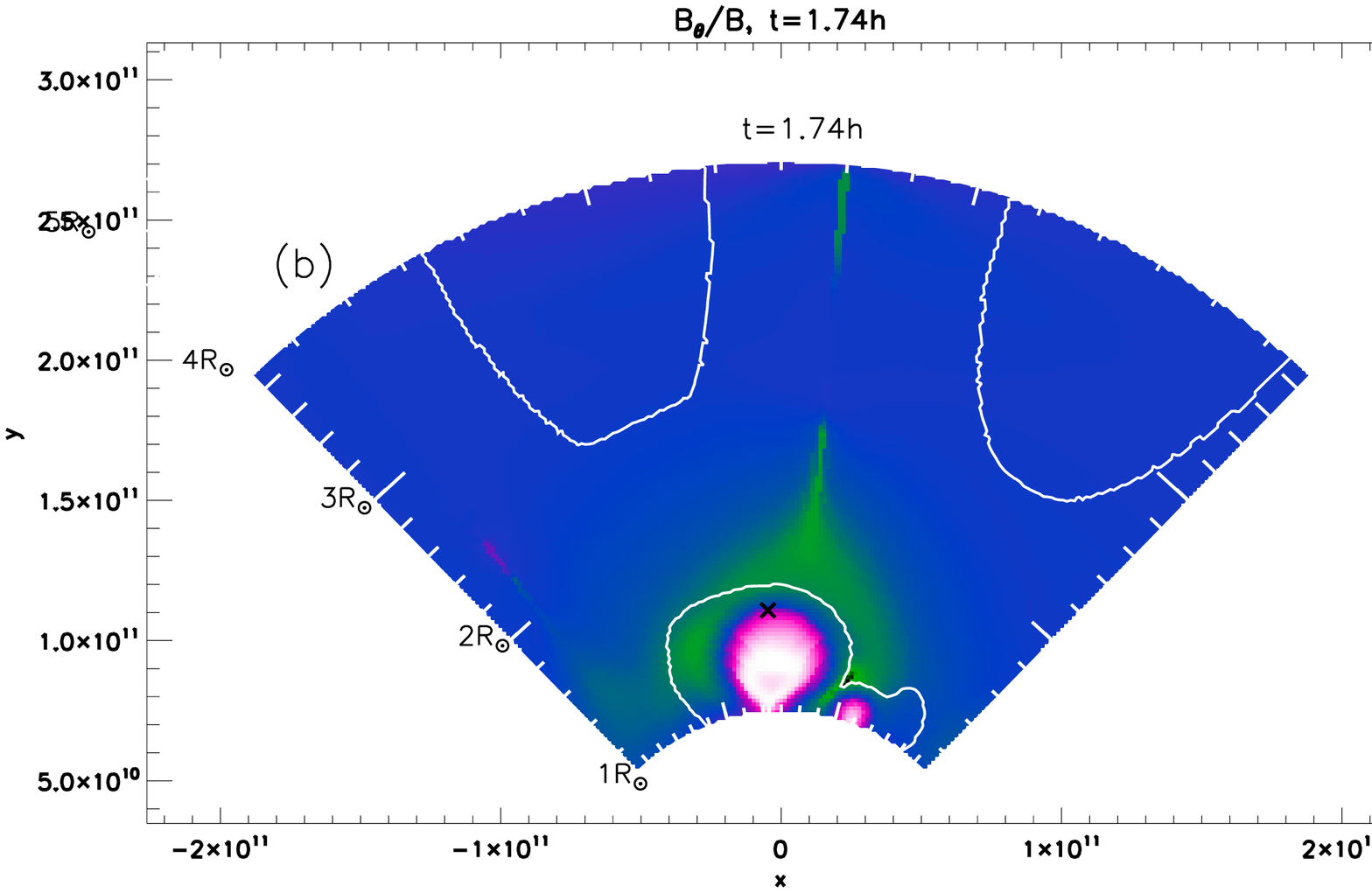}
\includegraphics[scale=0.28,clip,viewport=580 5 657 375]{Fig7b.ps}
\caption{Simulation with $T=2$ $MK$ and $B_{max}=7$ $G$.
Maps of $B_\theta/|B|$ in the $(r-\phi)$ plane
passing through the centre of the bipoles at $t=0$h and $t=1.74$h.
Superimposed is the contour line of $\beta=1$ (white line).
Black crosses indicate where the top of the flux rope is positioned.}
\label{evolT2B7}
\end{figure}

Following the location of the top of the flux rope we are able to track the ejection of the flux rope
and thus identify whether the flux rope ejection produces a CME
(i.e. extends beyond $r=4$ $R_{\odot}$)
or whether it is a quenched ejection.
We find that the simulations with $T=2$ $MK$ and $T=3$ $MK$ where $B_{max}=7$ $G$
do not show any CME propagation.
The flux rope only rises up to a given height and then stops.
In Fig. \ref{evolT2B7}, we show a quenched ejection for the simulation with $T=2$ $MK$ and $B_{max}=7$ $G$.
In this simulation the magnetic flux rope is stopped and remains positioned at $r=1.8$ $R_{\odot}$ after $1.74$h.

A key difference between the simulation where a CME develops (Fig.\ref{evolT2B21}a) and where the ejection is quenched (Fig. \ref{evolT2B7}a) is that
a high-$\beta$ region lies above the flux rope in the simulation where the ejection is quenched.
The fast CME can only develop in a low-$\beta$ environment.
In particular, the $\beta=1$ contour overlies the flux rope in Fig. \ref{evolT2B7}a,
while it only surrounds the null-point region in Fig. \ref{evolT2B21}a.
Although this is not a necessary or sufficient condition to determine the onset of a full or quenched ejection,
whether a high-$\beta$ region lies over the flux rope is one of the 
factors that can determine the CME evolution.

\subsubsection{Parameter space of CMEs}
\label{cmes}
\begin{figure}[!htcb]
\centering
\includegraphics[scale=0.42,clip,viewport=30 10 490 320]{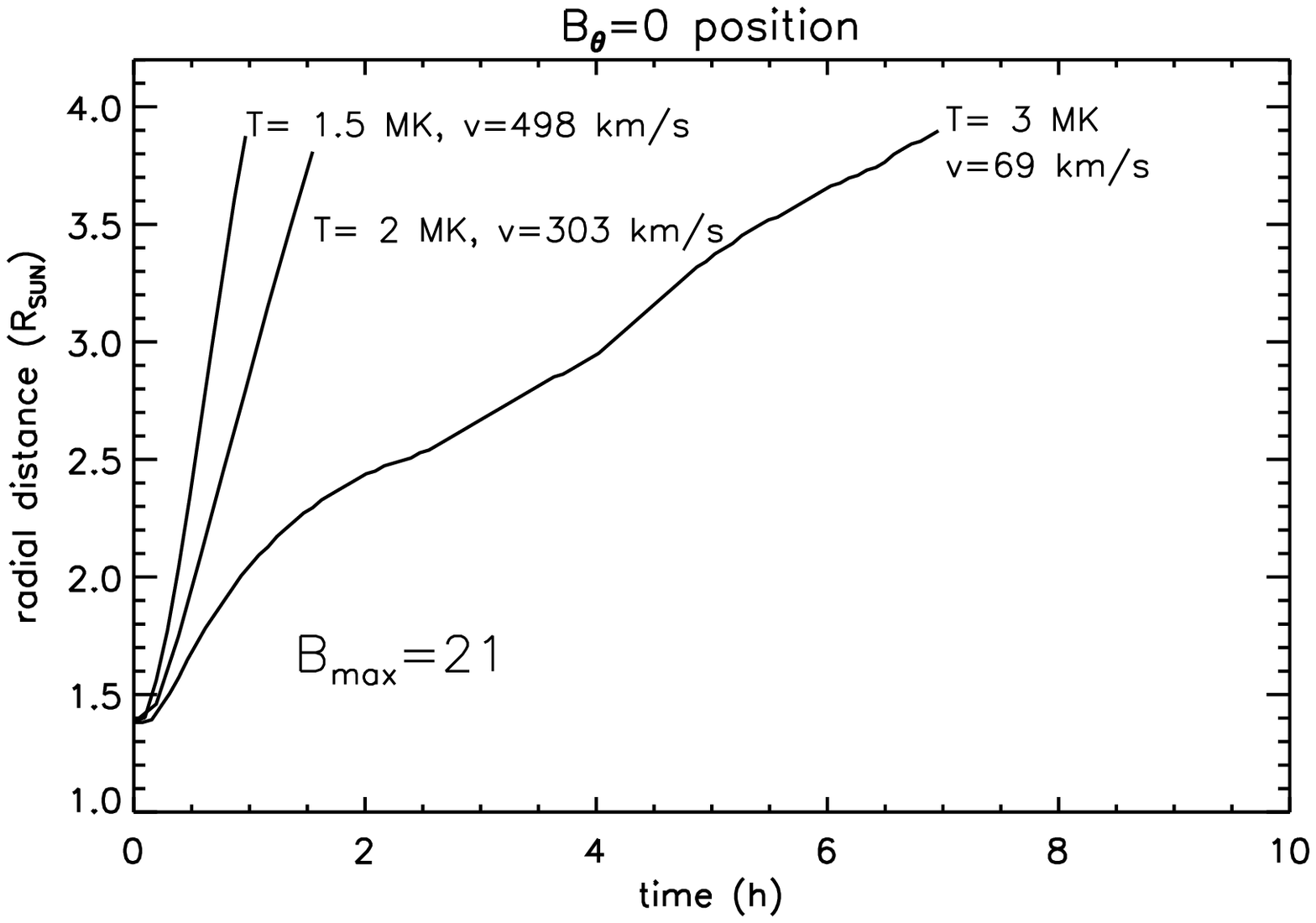}
\includegraphics[scale=0.42,clip,viewport=30 10 490 320]{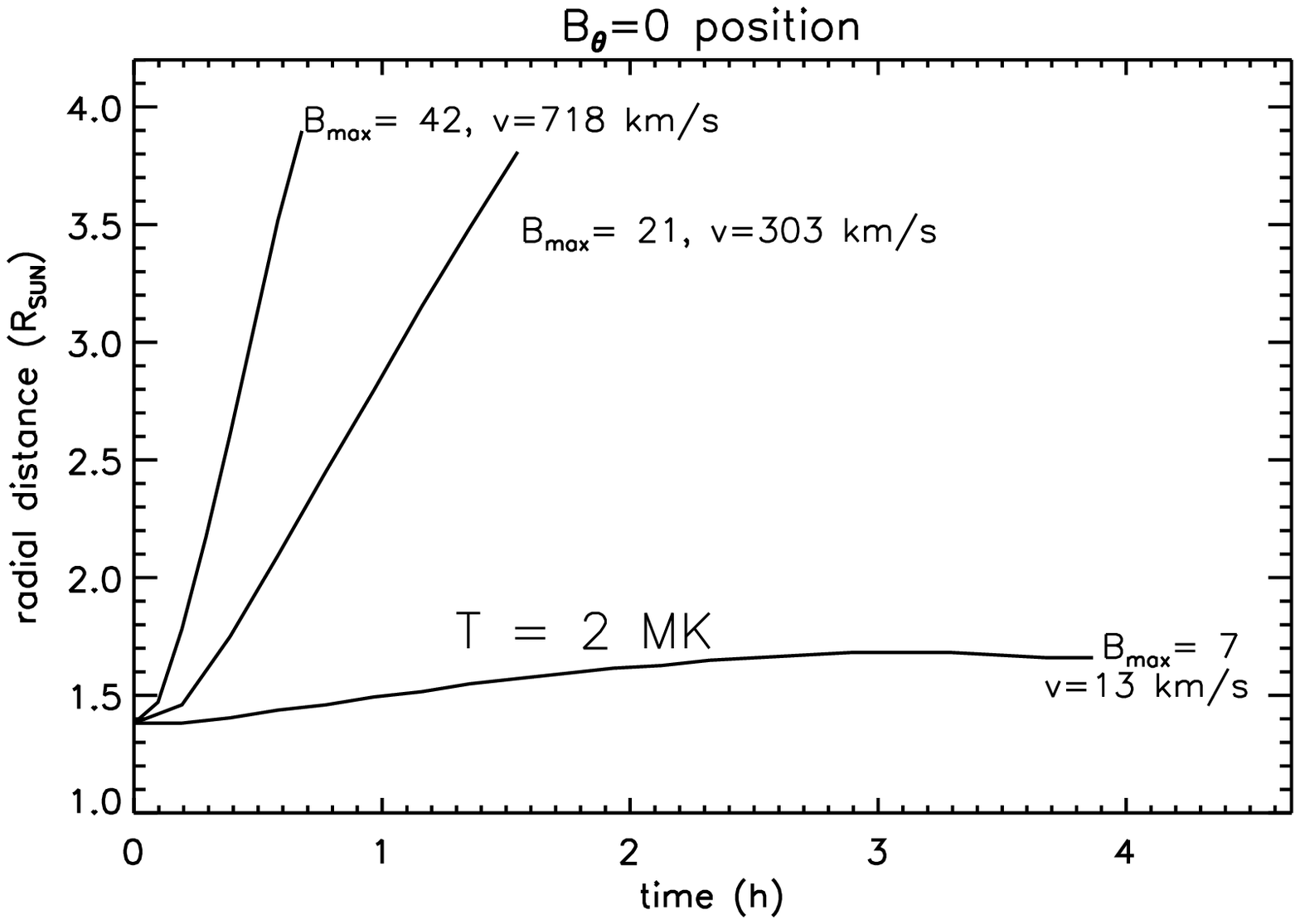}
\caption{(a) Position of the top of the flux rope as a function of time
in the three simulations with $B_{max}=21$ $G$.
(b) Position of the top of the flux rope as function of time
in the three simulations with $T=2$ $MK$.}
\label{velcme}
\end{figure}

All the remaining simulations show an ejection of the flux rope in a similar manner to the one in Sect. \ref{typicalsimulation}.
We do not present a detailed analysis here for each simulation,
but plot the position of the top of the flux rope as a function of time.
In Fig. \ref{velcme}a we show the position of the top of the flux rope for the three simulations where $B_{max}=21$ $G$
(central column in Fig.\ref{emptygrid}).
We stop tracking the top of the flux rope as soon as it is
close to the outer boundary and its propagation is affected by boundary effects.
The speed quoted in Fig. \ref{velcme} is the average speed of propagation.
The average speed of the CME spans a wide range from $69$ $km/s$ to $498$ $km/s$.
The lower the temperature, the faster the CME.
Also, the CMEs propagating in a $2$ $MK$ or $1.5$ $MK$ corona
are at near constant speed, while the speed varies slightly with radial distance for the slowest CME ($T=3$ $MK$).

Figure \ref{velcme}b shows the height of the top of the flux rope as a function of time for the three simulations with $T=2$ $MK$
(central row in Fig. \ref{emptygrid}).
The higher the magnetic field ($B_{max}$), the stronger the initial force due to the unbalanced Lorentz force under the 
magnetic flux rope, and the CME travels faster.
With $B_{\max}=7$ $G,$ we have a quenched ejection, while we have a $718$ $km/s$ fast CME with $B_{\max}=42$ $G$.
By changing the parameter $B_{max}$, we change the Alfv\'en speed in the system
and thus the evolution time scale.
It should be noted that the result of the simulations with different $B_{max}$
cannot be inferred solely through timescale arguments owing to the presence of the plasma.
For instance, this is demonstrated by the qualitatively different evolution of the simulation with $B_{\max}=7$ $G$
from the other simulations.

\begin{figure}[!htcb]
\centering
\includegraphics[scale=0.60,clip,viewport=120 80 430 270]{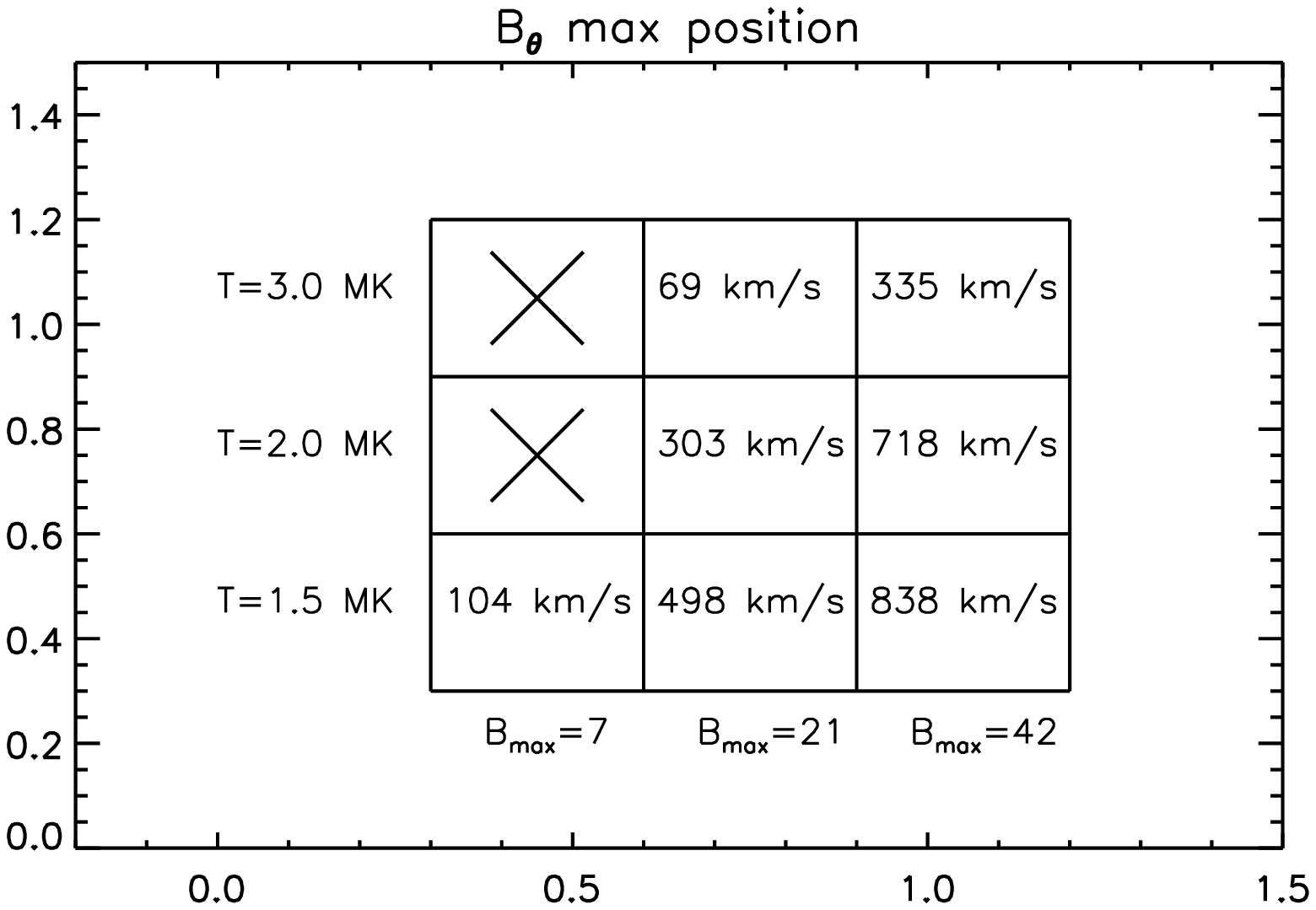}
\caption{Grid summarizing the average speed of the CME in the parameter space we investigate.
Crosses indicate a quenched ejection, and in the other boxes we write the value of the CME speed computed
from tracking the top of the flux rope as in Fig. \ref{velcme}.}
\label{velocitygrid}
\end{figure}

We note that the speed of the top of the flux rope is not necessarily the speed of the CME.
The top of the flux rope represents the leading edge of the CME, which
is subjected to the combined effect of
propagation and expansion. However, the results approximately indicate
the speed of the CME and its dependence on the parameters $T$ and $B_{max}$.
In Fig. \ref{velocitygrid} we summarize the speed of the CMEs for all of the simulations.
The fastest CME in our simulations can reproduce speeds of $838$ $km/s$,
and only two simulations show a quenched ejection.

\begin{figure}[!htcb]
\centering
\includegraphics[scale=0.42,clip,viewport=30 10 490 320]{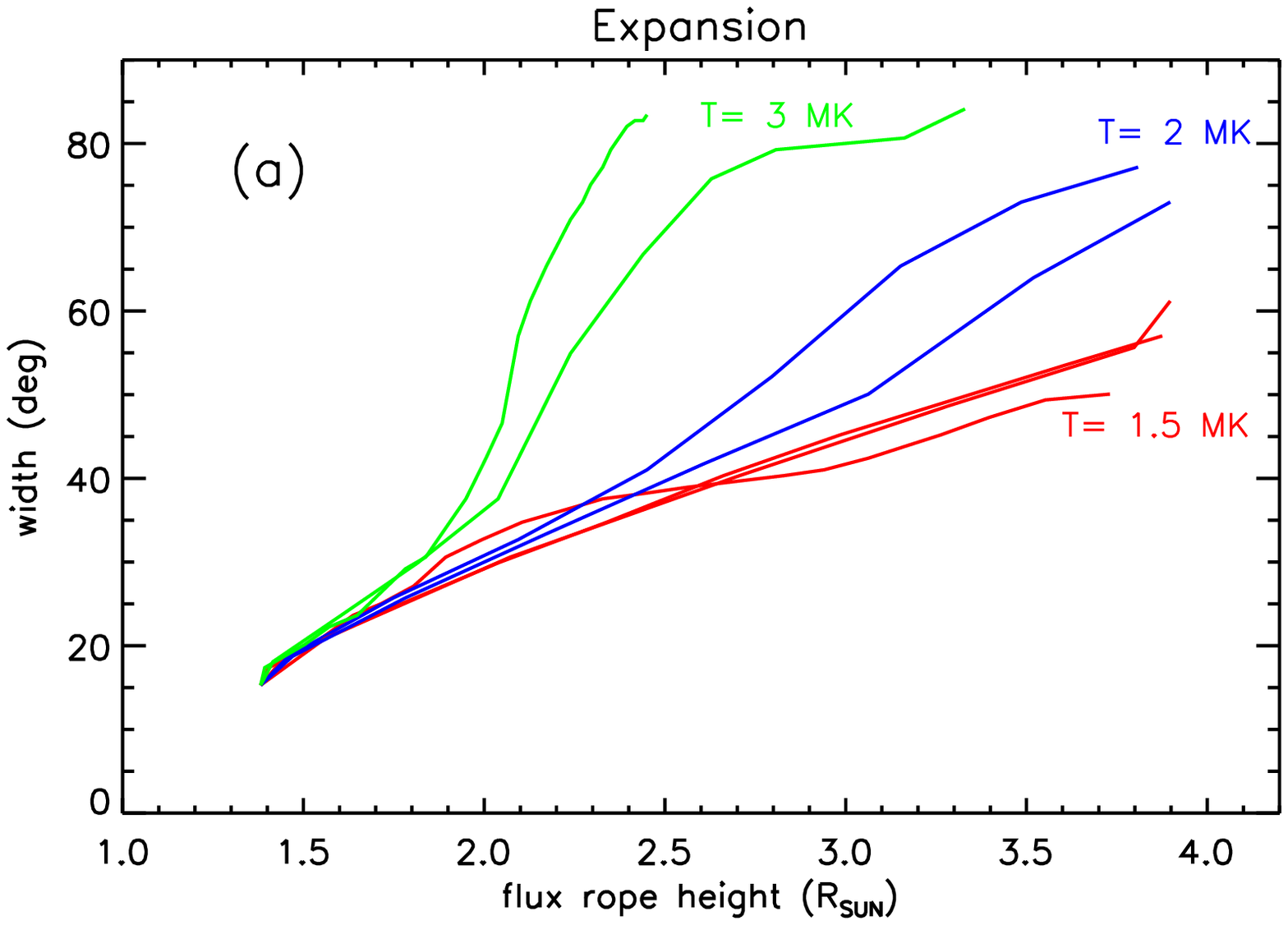}
\includegraphics[scale=0.33,clip,viewport=27 20 530 325]{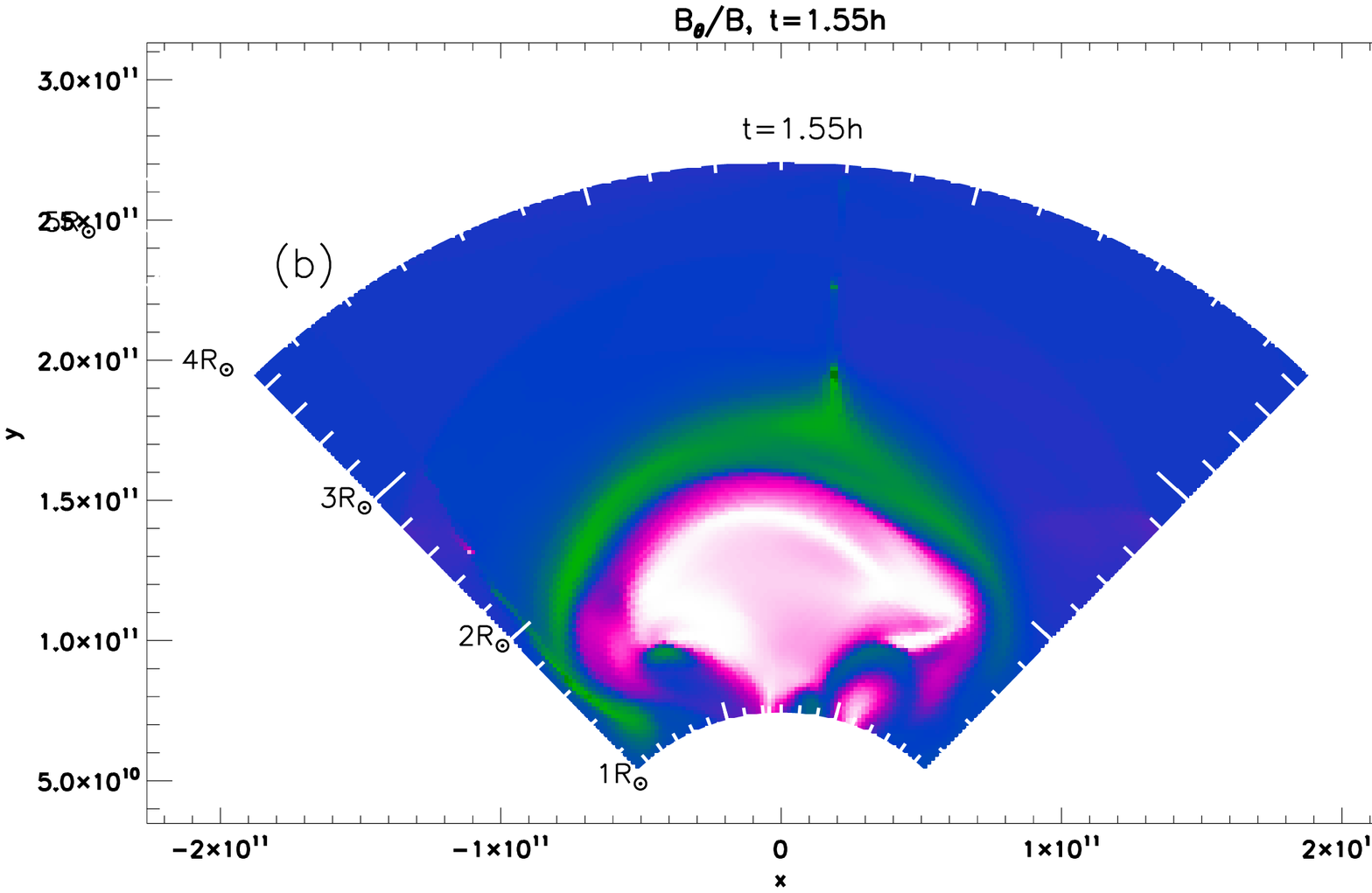}
\includegraphics[scale=0.28,clip,viewport=580 5 657 375]{Fig10b.ps}
\caption{(a) Angular extension of the flux rope as a function of the radial distance of the top of the flux rope
for all the simulations that show a CME. Simulations are grouped according to the simulation parameter $T$ given with different colours.
(b) Map of $B_\theta/|B|$ in the $(r-\phi)$ plane
passing through the centre of the bipoles at $t=1.55$h for the simulation with $T=3$ $MK$ and $B_{max}=21$ $G$.
The map show the full domain of our simulations from $r=1$ $R_{\odot}$ to $r=4$ $R_{\odot}$.}
\label{expansiongallt}
\end{figure}

We also use the quantity 
$B_{\theta}/|B|$ to follow the $\phi$ extension of the flux rope and thus its expansion.
We draw the contour of the flux rope by tracking where the cut of $B_{\theta}/|B|$
along the $\phi$ direction on the $ r$-$\phi$ plane passing through the centre of the bipoles
changes sign at different radial distances.
Then, we consider the maximum $\phi$ extension of the resulting flux rope contour.
Figure \ref{expansiongallt}a shows the extension of the flux rope
as a function of the height of the top of the flux rope for the simulations
where unquenched eruptions occur.
In this plot we group the simulations by temperature with different colours.
The intensity of the magnetic field $B_{max}$
does not significantly influence the expansion of the CME, i.e. its shape. 
This can be seen because lines with the same colour lie close to one another in Fig. \ref{expansiongallt}a.
In contrast, the background temperature, $T$, plays a significant role in shaping the ejection.

Below the height of $2$ $R_{\odot}$ the expansion proceeds in a similar manner for all the simulations.
However, around this radial distance the simulations with $T=3$ $MK$ show a larger expansion for higher radii.
In the simulations with $T=1.5$ $MK$ and $T=2$ $MK$, the expansion of the flux rope
proceeds nearly uniformly until the flux rope touches the upper boundary
where the extension is approximately $50^{\circ}$ and $70^{\circ}$, respectively.
In contrast, in the simulations with $T=3$ $MK$, the expansion rate increases at about $2$ $R_{\odot}$,
and the flux ropes covers an angle that is close to $80^{\circ}$ beyond $3$ $R_{\odot}$.
At the height of $2.5$ $R_{\odot}$, the simulations with $T=3$ $MK$ show an expansion that is
about twice as large as the expansion found in the simulations with a cooler stratified solar corona. 

In summary, Fig. \ref{expansiongallt}a shows that higher temperatures favour
higher expansion of the CME, which in turn results in a flattened shape.
The higher pressure and density above the ejecting flux rope
slow down the radial propagation and favour the expansion of the CME towards the sides.
An example of the different shape of the CME is given in Fig. \ref{expansiongallt}b
where we show a snapshot of the simulation with $T=3$ $MK$ and $B_{max}=21$ $G$
(to be compared with Fig. \ref{evolT2B21}e).

\section{Discussion}
\label{discussion}
In this paper we have considered the role of gravitational stratification 
on the propagation of a CME.
In order to do so, we investigated the parameter space by
tuning the temperature of the corona (i.e. gravity stratification)
and the maximum value of the magnetic field (i.e. the entire $\beta$ of the corona).
The techniques and the initial magnetic configurations
are identical to those of \citet{Pagano2013} where
a CME initiation was successfully modelled.

\subsection{Comparison with the flux rope ejection simulated in \citet{Pagano2013}}
The purpose of the present work is not to analyse the mechanism
that initiates the flux rope ejection in detail,
since it does not show any significant difference from what is
described in \citet{Pagano2013}.
However, once the onset of the ejection occurs, some differences
appear in the CME propagation
depending on the characteristics of the background corona, and
in the present paper we try to get some physical insight into the underlying physics involved.
In contrast to the background coronal model in \citet{Pagano2013},
the plasma pressure and density profile in the present simulations are
a consequence of gravitational stratification.

A drop in thermal pressure implies that the ejection encounters less resistance from the solar corona,
which should aid the production of a CME.
While this is true, it should be noted that
the ejected plasma is now subject to the gravity force
that was not considered previously.
The two effects nearly balance out since the speed of the ejection is generally comparable to 
the ones measured in \citet{Pagano2013}.
It should be noted, however, that slightly higher speeds are obtained in the present work.
Another aspect is that because of the lower thermal pressure in the outer solar corona,
the interaction between the ejected flux rope and  
the surrounding plasma is reduced, and 
this leads to a clearer three-component structure of the CME.
Our present simulations clearly show the light-bulb structure that was not seen in \citet{Pagano2013}.
For the same reason, significantly less magnetic reconnection due to numerical diffusion occurs
between the flux rope magnetic field and the external arcade magnetic field,
resulting in higher coherence of the expelled flux rope and less mixing between the flux- rope magnetic field lines 
and the external arcade magnetic field lines.

\subsection{Role of plasma $\beta$ on CME speed}
\textcolor{green}{\textcolor[rgb]{0,0,0}{In this study, the plasma $\beta$ plays a ma}}jor role in determining the evolution of the system.
Firstly, the plasma $\beta$ distribution determines
whether an ejection is quenched or if it will escape the solar corona.
In simulations where only low $\beta$ plasma covers the initial flux rope,
it quickly reaches a height of $4$ $R_{\odot}$.
In contrast to the two simulations (Fig. \ref{velocitygrid}) where the ejection stops,
the flux rope is surrounded by high $\beta$ plasma (see Fig. \ref{evolT2B7}a).
In our study we have used coronal values for magnetic field intensity, density of plasma, and temperature,
and our simulations seem to confirm that the onset of solar eruptions is caused by a purely magnetic process,
but the evolution of a flux rope ejection does depend on the plasma parameters.
This is an important factor to be considered when we wish to address the issue of CME generation,
especially in the space weather context.

Secondly, a lower plasma $\beta$ seems to favour higher CME speed.
In our investigation of the parameter space this was done by 
increasing the intensity of the flux rope magnetic field by a factor of 3 or 6.
Such an increase leads to an increase of a factor of 5 or 8 in the CME speed (simulations with $T=1.5$ $MK$ in Fig. \ref{velocitygrid}).
Similar results are found when comparing the CME speed in the ejecting simulations at $T=2$ $MK$ and $T=3$ $MK$,

More generally, Fig. \ref{velcme}b shows that we simulate two main evolution branches:
some simulations show a CME and others show a quenched ejection.
The plasma $\beta$ in the outer corona seems to be the parameter that determines
into which branch the system evolves (see Fig. \ref{profiles}c).
Of course, with a continuous variation of the physical properties of the system,
we expect intermediate cases that show a mixture of the two regimes.
This occurs when the plasma $\beta\sim1$ in the outer corona.
An example of such a regime is the simulation with $T=3$ $MK$ and $B_{max}=21$ $G$
where an extremely slow CME ($69$ $km/s$) does not travel at a constant speed (Fig. \ref{velcme}a).
In this simulation, high plasma $\beta$ lies above the flux rope, but at greater radial distance than in the other simulations where
the ejection stops. Similar behaviour was found in \citet{Pagano2013}.

\subsection{Role of gravitational stratification on CME speed}
In the framework of the present study, another way of tuning the plasma $\beta$ in the outer corona
is to change the coronal temperature.
In particular, the colder the atmosphere, the lower the plasma $\beta,$
and the more likely a full ejection occurs.
However, tuning the temperature only affects the capacity of the outer corona to react to the ejection,
but not the initial force to which the flux rope is subject.
For this reason, in our framework the temperature of the corona has a weaker effect compared to
the magnetic field intensity in producing CMEs at higher speeds.
At the same time, the role of the temperature stratification deserves 
more detailed analysis in future studies.
In fact, it is remarkable that simulations where the coronal temperature only differs by $0.5$ $MK$
show such a different behaviour. This could be the reason for the
enhancement of some CMEs in the solar corona.
Similar to the work of \citet{Lin2004b}, we find that the gravitational force
can be effective on the flux rope ejection when the magnetic field is relatively weak.
In particular, it can only slow down the ejected flux rope when the magnetic field is relatively weak.
In future, we need to investigate the cause of the flux rope ejections more.
In our study, the ejection is caused by the build up of an excess of the Lorentz force;
however, the question remains open whether a MHD instability,
like the Torus instability \citep{KliemTorok2006}, could explain the ejection.
In our case, a treatment like the one of \citet{KliemTorok2006} should include
the role of thermal pressure and gravity.
In doing so, the critical index for the enhancement of the ejection could be
determined as could whether it differs from previous studies.

\subsection{Role of gravity stratification on CME expansion}
Another result of our study is that the gravitational stratification can be responsible
for shaping the CME, since the thermal pressure ahead of the CME front can stimulate an expansion
at its sides.
We identified this effect in particular in the simulation with $T=3$ $MK$ and $B_{max}=21$ $G$
where the expanding CME acquires a shape that is significantly different from the ones found in other simulations.
However, the result is more general, as shown in Fig. \ref{expansiongallt}b.
From the analysis of our simulations, we find that the width of the CME bulb
can be significantly different from one simulation to the next if different coronal temperatures are considered.

Further studies are required, but in principle different stratifications may be responsible for the variety of CME shapes observed.
In particular, the CME we describe in the simulation with $T=3$ $MK$ and $B_{max}=21$ $G$ seems to be very similar to the ejection 
that occurred on June 13, 2010 as described by \citet{Gopalswamy2012}.
It is remarkable how the CME shape in our simulation matches the snapshot of the ejection at 5:40:54,
where not only is the front profile reproduced, but the northern side of the ejection also seems to behave
as if there is another bipole besides the one involved in the ejection as in our simulation (Fig. \ref{expansiongallt}b).
It should, however, be noted that in \citet{Gopalswamy2012} the CME is observed at $~1.4$ $R_{\odot}$, significantly lower than our snapshot in Fig. \ref{expansiongallt}b.
Also \citet{Patsourakos2010} studied the expansion of a CME and found the expansion to be
initially very rapid and characterized by a self-similar evolution at later times.
Our results differ from their findings,
since in our simulations the CME bubble expansion agrees with a constant decrease
in the aspect ratio (radial position/radius), and it never reaches the self-similar regime.
Such a difference in our results could be explained by the lack of an outflowing solar wind in our simulations that
would contribute to narrowing the propagation of the CME. 
This feature will be investigated in future studies.

\section{Conclusions}
\label{conclusion}

The present study aims at understanding the role of gravitational stratification in the 
propagation of CMEs.
To identify the role of gravitational stratification, we ran several simulations
that vary two quantities: the stratification temperature ($T$) and plasma $\beta$.
In all simulations an identical magnetic field configuration is used that has been proven to be suitable for an ejection. 

Our work appeared to produce a reasonable model for the early evolution of actual CMEs, since observations show that the CME acceleration tends to vanish above $4$ $R_{\odot}$ \citep{Vrsnak2001},
when the CME kinematics couple to the solar wind \citep{Gopalswamy2000b}.
Therefore our model covers the domain of initiation of a CME, and we can reasonably assume that 
all the ejections that leave our domain would not need any further acceleration mechanisms
to travel through interplanetary space dragged by the solar wind.

This study showed that gravitational stratification has an important effect on the propagation of CMEs in the solar corona
through the way it specifies how large the plasma $\beta$ becomes.
We also find that the plasma $\beta$ distribution is a crucial parameter that determines
whether a flux rope ejection escapes the solar corona, turning into a CME,
or if it just makes the flux rope find a new equilibrium at a greater height.
Similarly, we find that a cooler solar corona ($T\sim1.5$ $MK$) can help the escape of the CME and
make it travel faster.

Both of these results--first the importance of a low $\beta$ region above the magnetic flux rope to allow the ejection 
and secondly the role of the coronal temperature in the CME speed--can be tested with observations where
3D magnetic field reconstructions are carried along with simultaneous density and temperature diagnostics 
to infer the plasma stratification.
However, magnetic field reconstructions are more reliable from disk observations, which
makes it very difficult to infer coronal stratification because of line-of-sight integration effects.
However, future missions of the STEREO type with this capacity could be required.

\begin{acknowledgements}
The authors would like to thank Dr. Bernhard Kliem for useful discussions and suggestions.
The authors would also like to thank the referee for the constructive and positive feedback on the manuscript.
DHM would like to thank STFC, the Leverhulme Trust and the European Commission's Seventh Framework Programme
(FP7/2007-2013)  for their financial support.  PP would like to thank the European Commission's Seventh Framework Programme
(FP7/2007-2013) under grant agreement SWIFF (project 263340, www.swiff.eu) for  financial support.
These results were obtained in the framework of the projects
GOA/2009-009 (KU Leuven), G.0729.11 (FWO-Vlaanderen), and C~90347 (ESA Prodex 9).
The research leading to these results has also received funding from the European Commission's Seventh Framework Programme (FP7/2007-2013) under the grant agreements
SOLSPANET (project n 269299, www.solspanet.eu),
SPACECAST (project n 262468, fp7-spacecast.eu),
and eHeroes (project n 284461, www.eheroes.eu).
The computational work for this paper was carried out on the joint STFC and SFC (SRIF) funded cluster at the University of St Andrews (Scotland, UK).
\end{acknowledgements}

\appendix
\section{Equilibrium tests}
\label{appendix1}
To check the validity of our approach,
we performed some tests with a magnetic field configuration in force balance
and a stratified atmosphere.
In \citet{Pagano2013} a similar test was performed to verify that 
the Lorentz force is properly transported from the GNLFFF to the MHD model
without generating artificial forces.
We did not repeat a similar test, and we consider those results as valid.
Instead, we want to check that the 
stratified atmosphere and the extension of the domain to
$4$ $R_{\odot}$ does not introduce significant spurious forces
able to alter the stability of the system.

This test is performed using an initial force-free magnetic condition
similar to the case we wish to investigate,
consisting of two magnetic bipoles that lie close to one another but do not overlap.
We refer to \citet{Pagano2013} for further details on the magnetic configuration.

\begin{figure}[!htcb]
\centering
\includegraphics[scale=0.21,clip,viewport=27 20 530 325]{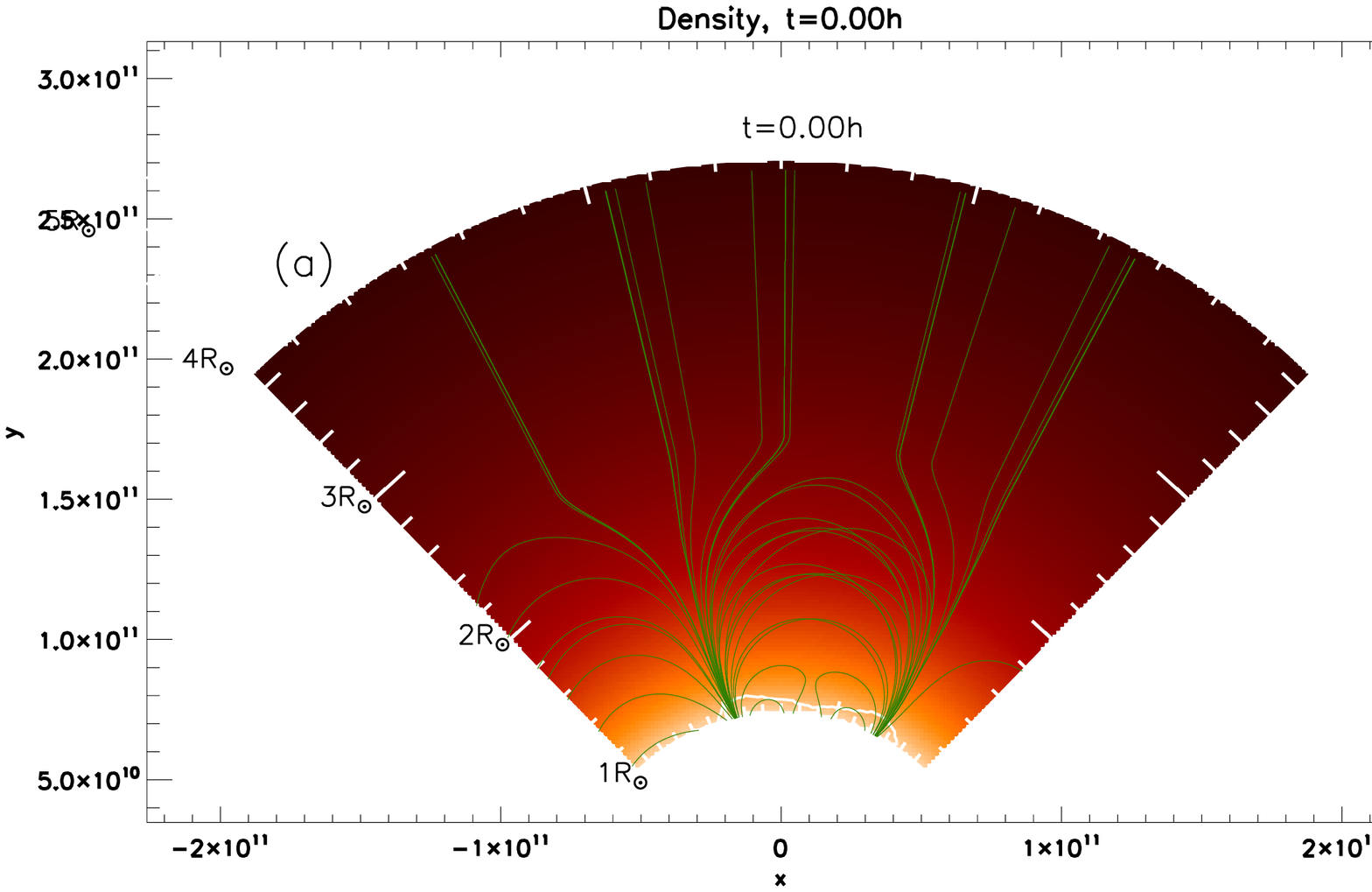}
\includegraphics[scale=0.18,clip,viewport=580 5 657 375]{FigA1a.ps}
\includegraphics[scale=0.21,clip,viewport=27 20 530 325]{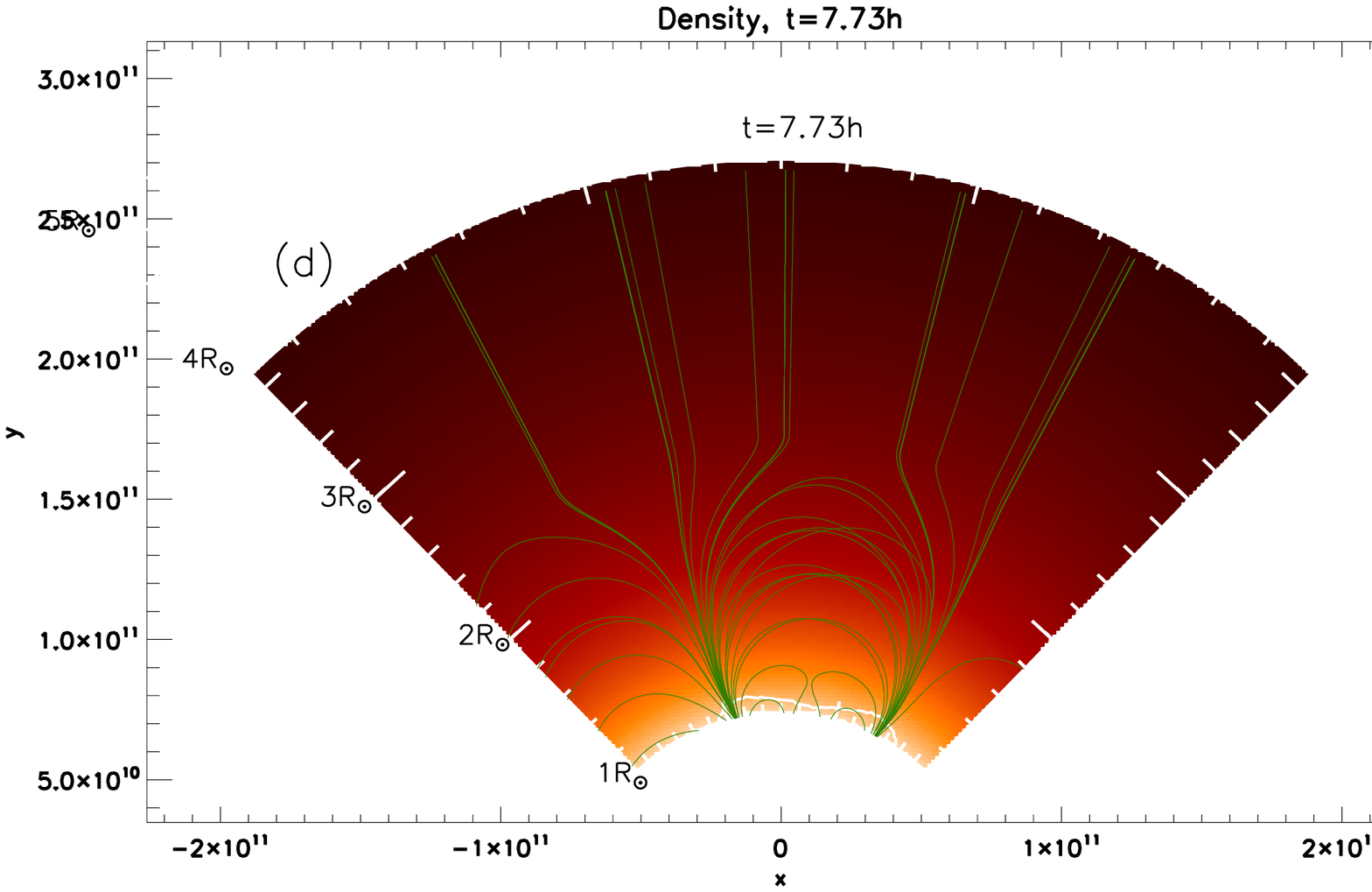}
\includegraphics[scale=0.18,clip,viewport=580 5 657 375]{FigA1d.ps}

\includegraphics[scale=0.21,clip,viewport=27 20 530 325]{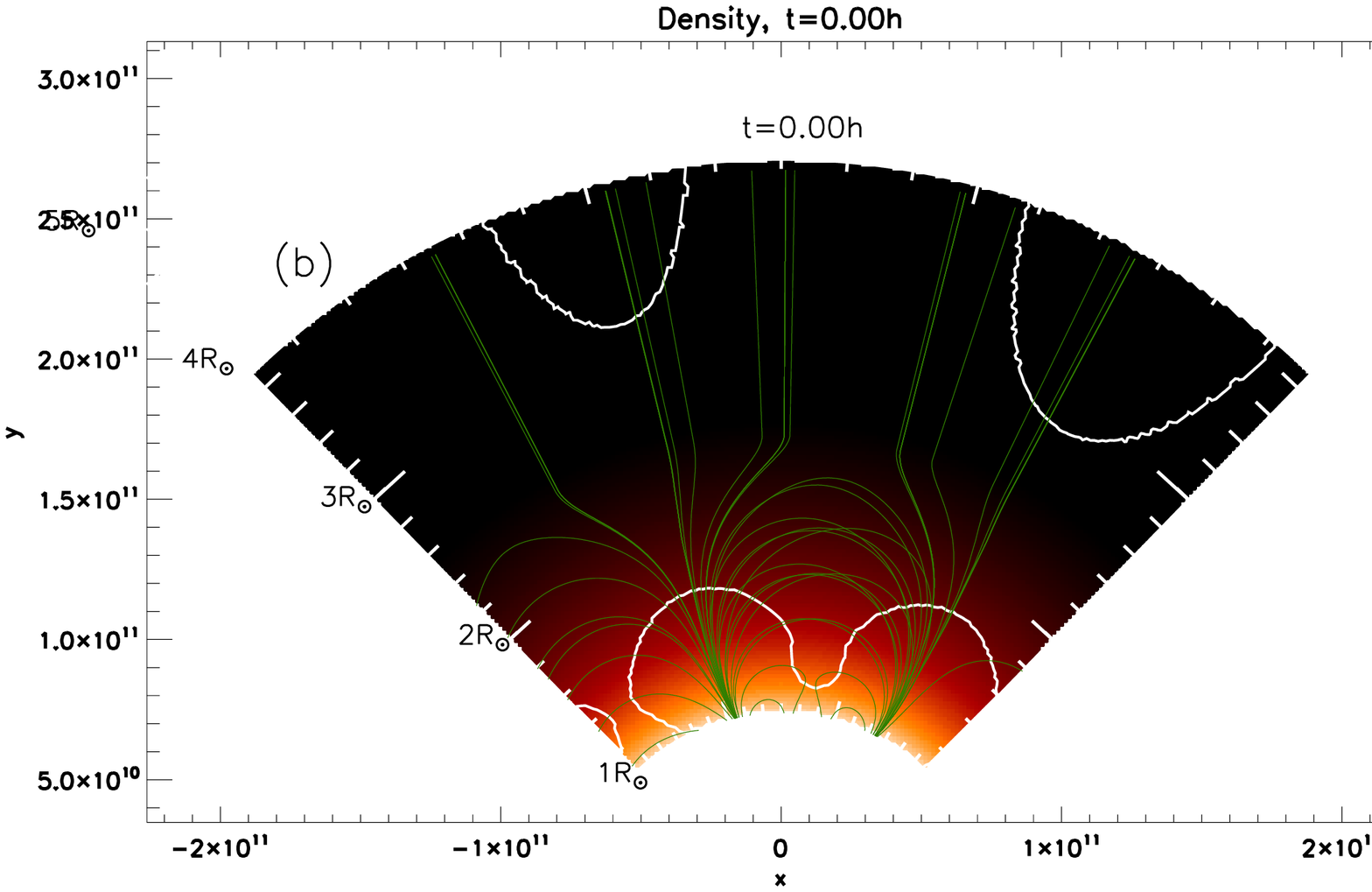}
\includegraphics[scale=0.18,clip,viewport=580 5 657 375]{FigA1b.ps}
\includegraphics[scale=0.21,clip,viewport=27 20 530 325]{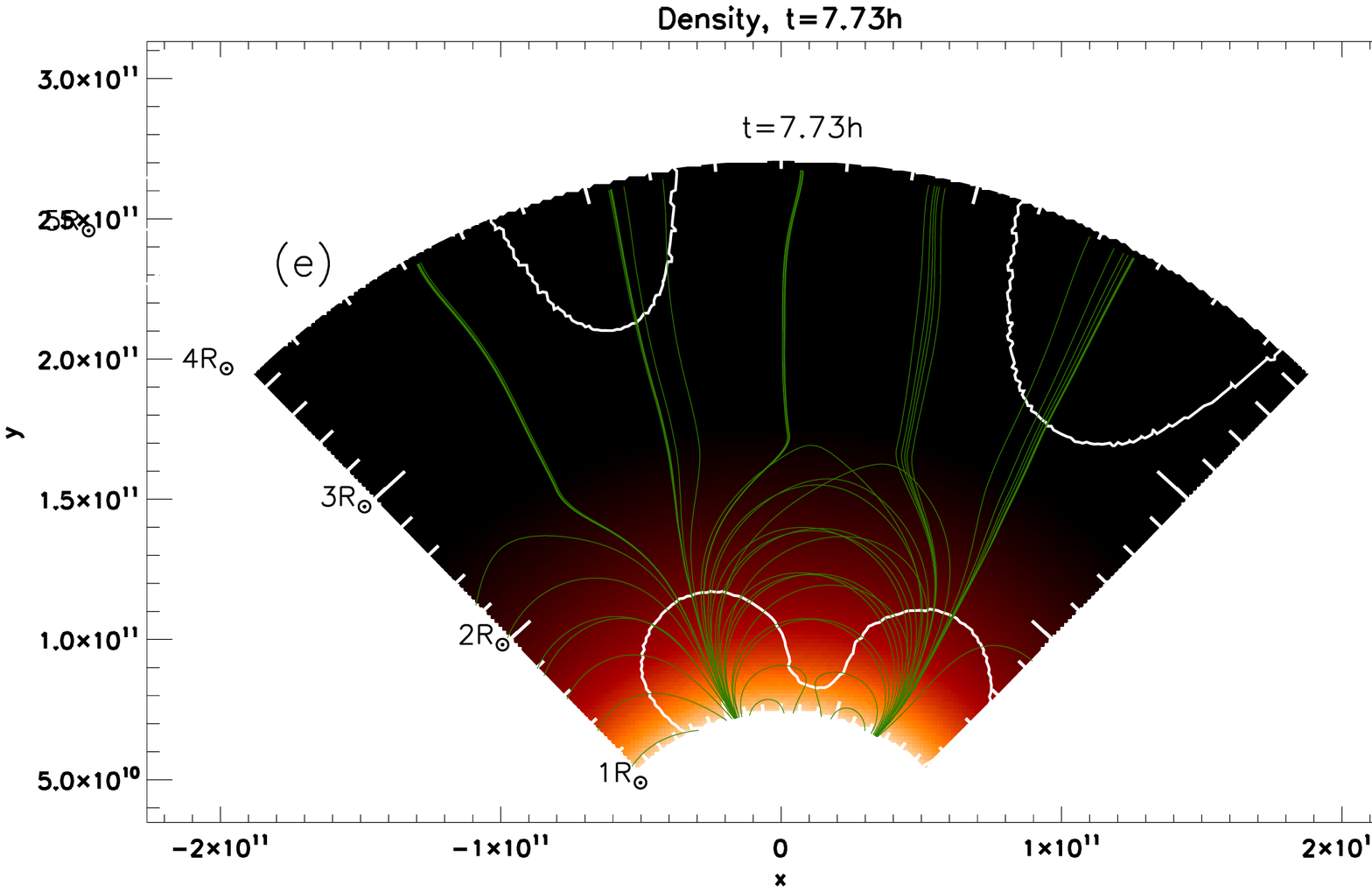}
\includegraphics[scale=0.18,clip,viewport=580 5 657 375]{FigA1e.ps}

\includegraphics[scale=0.21,clip,viewport=27 20 530 325]{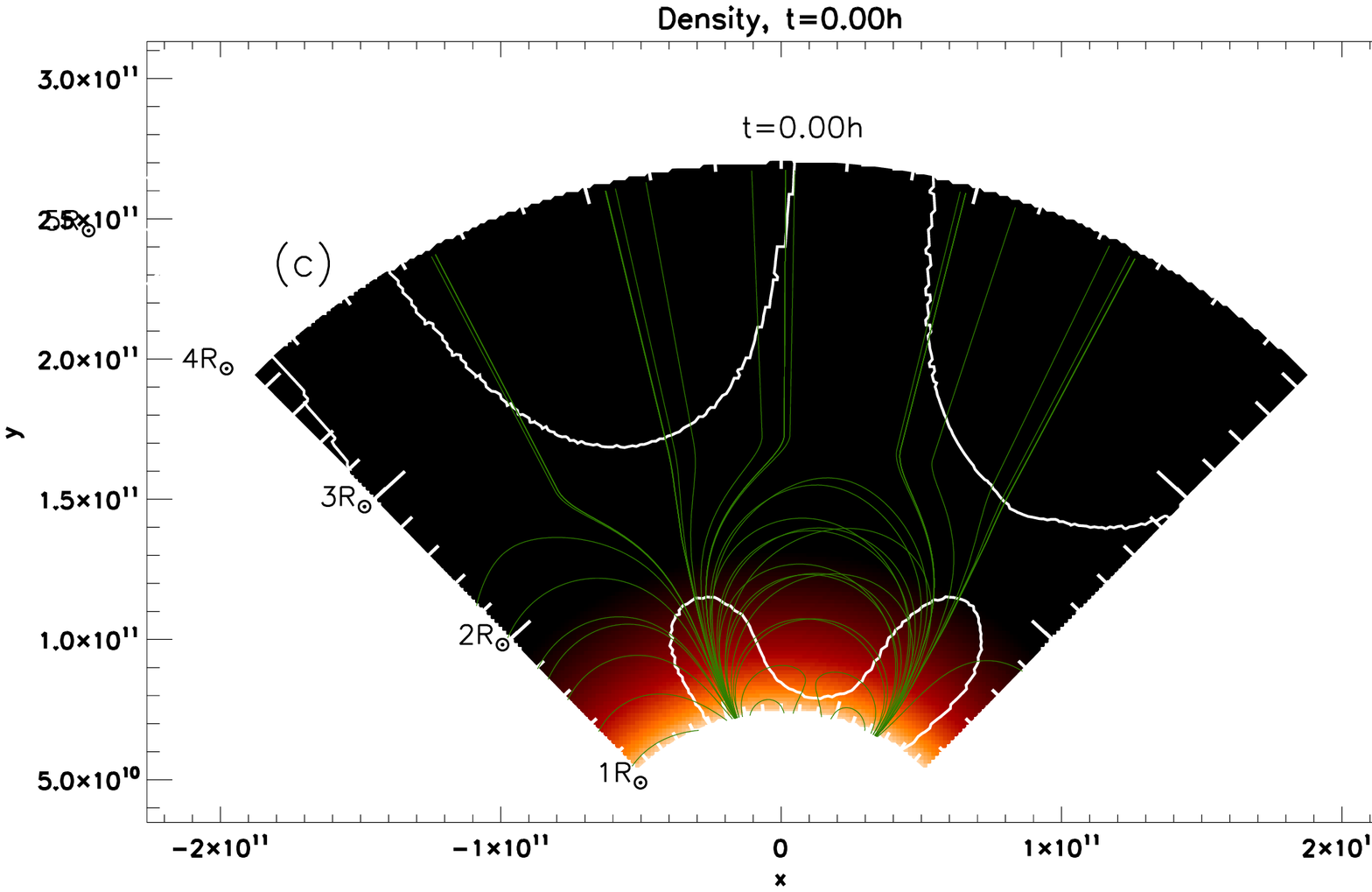}
\includegraphics[scale=0.18,clip,viewport=580 5 657 375]{FigA1c.ps}
\includegraphics[scale=0.21,clip,viewport=27 20 530 325]{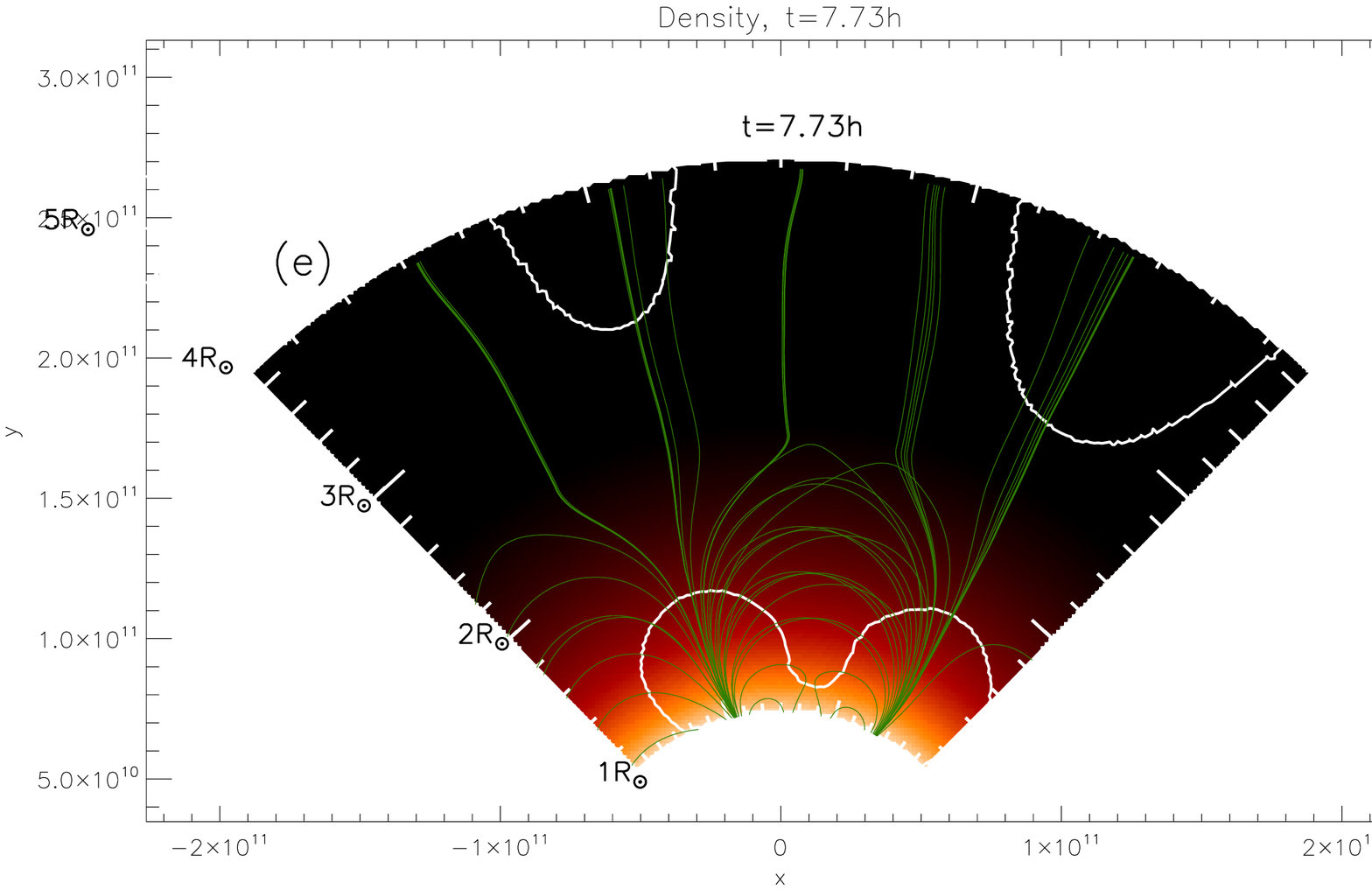}
\includegraphics[scale=0.18,clip,viewport=580 5 657 375]{FigA1f.ps}
\caption{Maps of $Log_{10}(\rho[g/cm^3])$ in the $(r-\phi)$ plane
passing through the centre of the bipoles at different times for different test simulations.
Superimposed are magnetic field lines plotted from the same starting points (green lines)
and the contour line of $\beta=1$ (white line).
(a) and (d) $T=3$ $MK$, $B_{max}=7$ $G$;
(b) and (e) $T=2$ $MK$, $B_{max}=21$ $G$;
(c) and (f) $T=1.5$ $MK$, $B_{max}=42$ $G$.}
\label{testeq}
\end{figure}

We consider several stratified density and pressure profiles that cover the parameter space
described in Sect. \ref{parameterspace}.
Namely, we show here in Fig.\ref{testeq} simulations with ($T=3$ $MK$, $B_{max}=7$ $G$), ($T=2$ $MK$, $B_{max}=21$ $G$), and ($T=1.5$ $MK$, $B_{max}=42$ $G$).

To test the stability, we let the system evolve for a large number of Alfv\'en times.
In the simulation with $B_{max}=7$ $G$, we estimate the Alfv\'en time $\tau_{Alf}=120$ $s$
as the time it takes an Alfv\'en wave to travel across a single bipole with a typical speed of $800$ $km/s$.
Consequently, the Alfv\'en time is $\tau_{Alf}=40$ $s$ and $\tau_{Alf}=20 s$ for the simulations
with $B_{max}=21$ $G$ and $B_{max}=42$ $G$. 
Fig.\ref{testeq} shows the density contrast (defined in Eq. \ref{densitycontrast}).
The plasma density contrast does not show any significant evolution
for any of the test simulations, even though small-scale changes occur.

The only test that shows an observable change in the magnetic structure is the one with
$T=1.5 MK$ and  $B_{max}=42$ $G$.
This is due to the non-zero values of $p_{bg}$ and $\rho_{bg}$ that lead to 
a downward bulk motion of plasma in the higher part of the corona,
as explained in Sect. \ref{parameterspace}.
After this occurs, numerical reconnection takes place, and 
some of the initially open magnetic field lines reconnect near the outer boundary.
However, only a small change in the magnetic configuration can be seen
after $\sim3.3  $h, equivalent in this particular simulation to $600$ $\tau_{alf}$ .
In the corresponding simulation in Sect. \ref{spaceparamaterCME}
where a CME occurs,
the magnetic flux rope reaches the top of the domain in less than $0.8$h.
Therefore the motions due to the non-zero values of $p_{bg}$ and $\rho_{bg}$
are negligible in comparison to the dynamics produced by the flux rope ejection.

These tests show that the transporting between the GNLFFF and MHD models is 
possible when including the effects of
gravity, a stratified atmosphere, and magnetic field intensity tuning
and that these changes 
do not generate numerical artefacts in the MHD simulation.

\bibliographystyle{aa}
\bibliography{ref}

\end{document}